\documentclass[11pt]{article}

\usepackage[a4paper,margin=0.72in]{geometry}
\usepackage{graphicx}
\usepackage{amsmath,amssymb}
\usepackage{booktabs}
\usepackage{longtable}
\usepackage{multirow}
\usepackage{array}
\usepackage{threeparttable}
\usepackage{siunitx}
\usepackage{caption}
\usepackage{enumitem}
\usepackage{url}
\usepackage{hyperref}
\usepackage{orcidlink}
\usepackage{microtype}
\usepackage{xurl}
\usepackage{placeins}
\usepackage{float}

\begin{document}

\begin{center}
{\LARGE \bfseries A relational fabrication-to-modeling database for memristor devices\par}

\vspace{1.0em}

{\large
Lai Gan$^{1,*}$\,\orcidlink{0009-0004-9033-3390},
Guoyang Huang$^{1}$\,\orcidlink{0000-0002-0964-7367},
Deepika Yadav$^{1}$\,\orcidlink{0009-0000-2394-8329},
Spyros Stathopoulos$^{1}$\,\orcidlink{0000-0002-0833-6209},\\[0.3em]
Ben D. Rowlinson$^{1}$\,\orcidlink{0000-0002-7533-0579},
Themis Prodromakis$^{1,*}$\,\orcidlink{0000-0002-6267-6909}
\par}

\vspace{0.8em}

{\normalsize
$^{1}$ Centre for Electronics Frontiers, School of Engineering, University of Edinburgh, Edinburgh, UK\\[0.4em]
$^{*}$ Corresponding authors: \href{mailto:L.Gan-9@sms.ed.ac.uk}{L.Gan-9@sms.ed.ac.uk};
\href{mailto:t.prodromakis@ed.ac.uk}{t.prodromakis@ed.ac.uk}
\par}
\end{center}

\vspace{1.0em}

\section*{Abstract}

Resistive random access memory (RRAM) devices, also known as memristors, are highly dependent on fabrication, location on the wafer, and measurement protocols. However, openly available large-scale memristor datasets remain scarce, particularly those that combine experimental metadata with electrical measurements and preserve explicit links across the experimental workflow. Here, we present a comprehensive relational database of automated electrical characterization data from oxide-based memristors. The database links 6,190 memristor devices across TiN/HfO$_\text{x}$N$_\text{y}$/TiN, Pt/TiO$_\text{x}$/AlO$_\text{y}$/Pt, and Pt/TiO$_\text{x}$/Pt stacks to 161,006 validated experiments and over 169 million electrical point records. It integrates fabrication, wafer mapping, electrical characterization, and modeling to describe electroforming, current-voltage non-linearity, memory windows (R$_{\text{off}}$/R$_{\text{on}}$), switching dynamics, and short-term volatility. Released as a normalized and indexed SQLite database with schema documentation, graphical user interfaces, and examples of empirical modeling, this resource supports provenance-aware querying, statistical analysis, and data-driven applications for the development of future memristor technologies.

\textbf{Keywords:} memristor; database structure; SQL; experimental data; resistive memory

\section*{Background \& Summary}
Resistive random access memory (RRAM) devices, or memristors, are two-terminal devices whose resistance can be modulated by electrical stimulation.  The memristor was originally introduced by Chua as a circuit element relating charge and magnetic flux \cite{Chua1971a}. This concept was later linked to nanoscale metal-oxide devices exhibiting history-dependent resistive switching and pinched current-voltage hysteresis \cite{Strukov2008,Waser2007Nanoionics,Wong2012}. The device features have made them a promising candidate for non-volatile memory, in-memory computing, neuromorphic hardware, and reconfigurable electronics \cite{Ielmini2018b,Lanza2022a,Aguirre2024}.

Despite their simple two-terminal geometry, memristor measurements are highly context-dependent. The observed response depends on the material stack, device geometry, pristine-state condition, electroforming or preconditioning history, compliance settings, voltage-sweep direction, pulse amplitude, pulse width, read bias, and prior resistance state. In oxide-based systems, resistive switching is commonly associated with coupled ionic, electronic, interfacial, and filamentary processes, which can contribute to device-to-device and cycle-to-cycle variability \cite{Waser2007Nanoionics,Yang2018c,Dittmann2021a,Li2017CFDynamics,Jiang2017Stochasticity}. Methodological work has therefore emphasized the importance of reporting fabrication, variability, yield, and statistically informative summaries \cite{Lanza2019RecommendedMethods,Lanza2021a}. These considerations motivate data structures that preserve explicit links across the process-device-measurement-feature workflow, particularly for automated electrical characterization.

Here, we introduce a large-scale relational database for oxide-based memristor devices, built from an automated measurement workflow adapted from a prior benchmarking methodology~\cite{Stathopoulos2019}. The database links fabrication, wafer mapping, electrical characterization, and modeling, providing a provenance-aware, AI-ready resource for statistical and data-driven analysis and supporting upstream device-technology optimization and downstream circuit and system design applications.

\section*{Methods}
\label{sec:Methods}

\begin{figure}[H]
    \centering
    \includegraphics[width=0.8\linewidth]{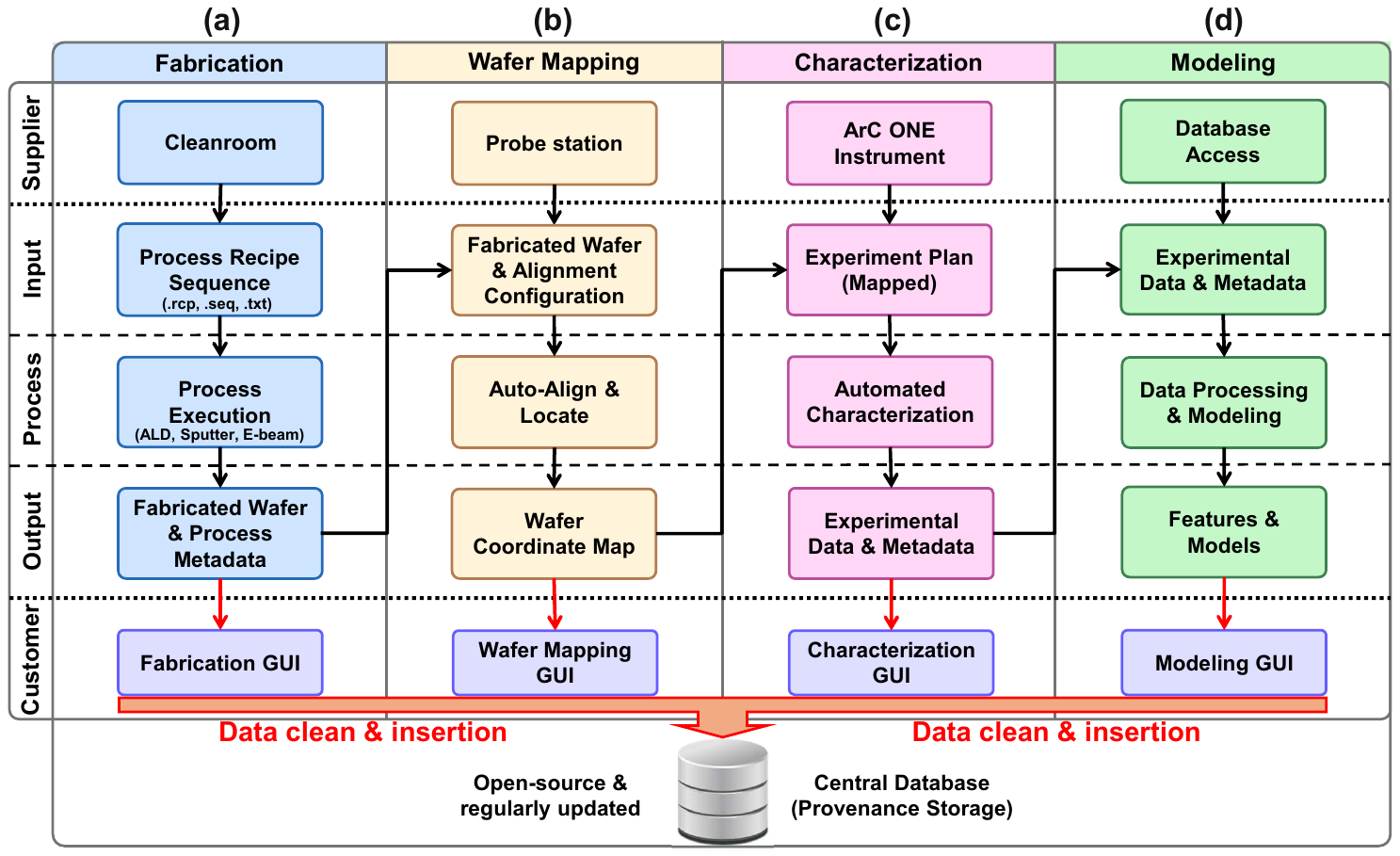}
    \caption{\textbf{
    Memristor data processing experimental pipeline.
}
SIPOC framework summarizing the four workflow stages: (a) Memristor Fabrication, (b) Wafer Mapping, (c) Electrical Characterization, and (d) Modeling, with curated records linked to a central provenance-aware database.
}
    \label{fig:data_measure_procedure}
\end{figure}

This section describes how the dataset was generated, organized, and curated across the four-stage experimental workflow: fabrication, wafer mapping, electrical characterization, and modeling. This workflow is summarized in Figure~\ref{fig:data_measure_procedure} using a supplier-input-process-output-customer (SIPOC) framework.

\label{subsec:Methods_Memristor}

\phantomsection
\label{subsec:methods_fabrication}
\noindent\textbf{Memristor fabrication.}
The first stage of the memristor data pipeline is the fabrication workflow summarized in Figure~\ref{fig:data_measure_procedure}(a). The memristor devices in this study were fabricated as vertical metal-insulator-metal (MIM) stacks using identical mask designs. Three material stacks are implemented: TiN/HfO$_\text{x}$N$_\text{y}$/TiN, Pt/TiO$_\text{x}$/AlO$_\text{y}$/Pt, and Pt/TiO$_\text{x}$/Pt. TiN electrodes are deposited via reactive sputtering of a Ti target in an Ar/N atmosphere. Pt electrodes are deposited by e-beam evaporation. TiO$_\text{x}$/AlO$_\text{y}$ bi-layer insulators are deposited by reactive sputtering of a Ti target in an Ar/O atmosphere, followed by reactive sputtering of an Al target in an Ar/O atmosphere. HfO$_\text{x}$N$_\text{y}$ is deposited using a super-cycled atomic layer deposition (ALD) process with alternating H$_2$O, N$_2$, Ar, and tetrakis(dimethylamino)hafnium (TDMAHf) precursor pulses. Supplementary Section~\ref{supp-sec:fabrication} summarizes the fabrication parameters and the recipes recorded directly from the fabrication tools. The associated recipe construction user interface is shown in Figure~\ref{fig:recipe_construction_gui}, Figure~\ref{supp:file_attachment_interface}, and Figure~\ref{supp:ald_cycle_editor}.

\phantomsection
\label{subsec:methods_wafer_mapping}
\noindent\textbf{Wafer mapping.}
The second stage of the memristor data pipeline is the wafer mapping workflow, summarized in Figure~\ref{fig:data_measure_procedure}(b). This stage establishes how memristor device locations on the wafer are recorded and indexed in the database. Each 150\,mm Si substrate wafer is organized into 192 square dies, each die into 9 sub-dies, and each sub-die into 32 discrete memristor devices. The 32 discrete memristors are indexed by the wordline and bitline coordinates. Each electrical measurement experiment can therefore be linked to a unique physical device location index, using wordline-bitline addressing within the Stand-Alone layout considered in this dataset (SA in the database).\cite{Lanza2019RecommendedMethods,Papandroulidakis2019MCMTLG}

During automated acquisition, wafer positioning was controlled by the MPI SENTIO probe station via a custom interface built on the SENTIO Python library \cite{SentioAPIDoc}. Figure~\ref{fig:wafer_heatmap}(a) plots the experimental counts at the wafer, die, and sub-die levels. Device selection and electrical characterization were then performed using the ArC ONE instrument with a 32\,$\times$\,32 probe card, controlled by custom Python scripts adapted from the vendor documentation \cite{ArcOneProduct} and the open-source \texttt{arc1\_pyqt} repository \cite{arc1_pyqt}. Figure~\ref{fig:wafer_heatmap}(b) shows the experiment counts within a sub-die addressed through the ArC ONE board.

\phantomsection
\label{subsec:methods_electrical_characterization}
\noindent\textbf{Electrical characterization.}
The third stage of the memristor data pipeline is the electrical characterization workflow summarized in Figure~\ref{fig:data_measure_procedure}(c). For each memristor device, at least one complete standard characterization set was implemented, with interrupted experiments remeasured to complete the set. Following the benchmarking methodology~\cite{Stathopoulos2019}, our measurement protocol comprises 3 measurement blocks and 20 automated experiments. The first measurement block uses 10 experiments for electroforming feature extraction (Figure~\ref{fig:ef_data}). The second measurement block uses 6 experiments for characterizing current-voltage (I-V) non-linearity and memory window (R$_{\text{off}}$/R$_{\text{on}}$) (Figure~\ref{fig:hys_data} and Figure~\ref{fig:ronroff_data}(b)). The third measurement block uses 4 experiments for characterizing switching dynamics and volatility (Figure~\ref{fig:pf_data} and Figure~\ref{fig:volatility_data}). Further protocol details are provided in Supplementary Section~\ref{supp-sec:Measurement protocol}.

In the first measurement block, the 10 experiments used for electroforming feature extraction were implemented using the \texttt{CurveTracer} (\texttt{CT}) function available in ArC ONE \cite{ArcOneProduct,ArcOneUserGuide,arc1_pyqt}. Across these \texttt{CT} experiments, symmetric positive and negative voltage sweeps were applied, with the maximum sweep voltage increased from 3.0 to 7.5\,V in 0.5\,V steps. A compliance current limit of $\pm 10\,\si{\micro\ampere}$ was applied throughout this block to reduce the risk of irreversible breakdown during electroforming (Supplementary Section~\ref{supp-sec:Measurement block one}). In the second measurement block, the 6 post-forming experiments also used the \texttt{CT} function. Across these experiments, symmetric positive and negative voltage sweeps were applied, with the maximum sweep voltage increased from 0.5 to 3.0\,V in 0.5\,V steps. The compliance current limitation was removed in this block to capture the post-forming I-V hysteresis response unconstrained by current compliance, from which I-V non-linearity and R$_{\text{off}}$/R$_{\text{on}}$ were extracted (Supplementary Section~\ref{supp-sec:Measurement block two}). In the third measurement block, the 4 experiments used the \texttt{ParameterFit} (\texttt{PF}) and \texttt{ParameterFit\_interRetention} (\texttt{PF-IR}) functions to characterize switching dynamics and volatility. These experiments were performed at pulse widths of 10, 50, 100, and 500\,\si{\micro\second} (Supplementary Section~\ref{supp-sec:Measurement block three}). In \texttt{PF-IR}, each programming segment was followed by a non-invasive read segment, allowing both programming-induced resistance changes and post-programming resistance relaxation to be tracked.

\phantomsection
\label{subsec:methods_modeling}
\noindent\textbf{Memristor feature extraction and modeling.}
The fourth stage of the memristor data pipeline is the modeling workflow summarized in Figure~\ref{fig:data_measure_procedure}(d). Using the validated electrical characterization data from the Electrical Characterization stage, five feature groups were derived to describe device behavior.

The first and second measurement blocks both used the \texttt{CT} function, for electroforming and post-forming I-V characterization. Each \texttt{CT} experiment was decomposed into four voltage-sweep branches: $0 \rightarrow V_{\max}$, $V_{\max} \rightarrow 0$, $0 \rightarrow V_{\min}$, and $V_{\min} \rightarrow 0$. The corresponding applied-voltage sequences are shown in Figure~\ref{fig:ef_data}(a) for the electroforming block and Figure~\ref{fig:hys_data}(a) for the post-forming block. Prior studies have used $\sinh$ functions to describe non-linear I-V behavior, including responses associated with electron tunneling and related transport processes \cite{Yang2008Memristive,Yakopcic2013,Messaris2018}. Based on this precedent, each branch was fitted independently in the low-voltage window ($|V| \leq 0.5$\,V) using the following $\sinh$ function:
\begin{equation}
    I(V)=a\sinh(bV),
    \label{eqn:sinh_fit_common}
\end{equation}

\phantomsection
\label{subsubsec:ef_features}
\noindent\textbf{Feature Group 1: Electroforming Features.}
Given the range of material stacks and device dimensions in the dataset, electroforming was not defined using a single fixed voltage or resistance threshold. Prior work in the academic literature indicates that electroforming is stochastic, varies across devices \cite{Skaja2018,Yun2021}, and depends on material, dimensions, and oxide properties \cite{Leonetti2023,Zhang2022}. 

Electroforming was assessed for each branch within the low-voltage window ($|V| \leq 0.5$\,V), which was used to probe the existing memristor response rather than to intentionally drive further switching. Two indicators were used. First, a compliance-current hit within this window was treated as evidence that the branch had already entered a highly conductive state, consistent with the formation of a conductive pathway. Second, when no compliance-current hit was observed, the coefficient of determination ($R^2$) from the $\sinh$ fitting in Equation\,\ref{eqn:sinh_fit_common} was used to quantify agreement between the measured low-voltage I-V response and the fitted $\sinh$ response.

Branches with $R^2 \geq 0.95$ were assigned Code~1.0, indicating good agreement; branches with $0.5 \leq R^2 < 0.95$ were assigned Code~0.5, indicating uncertain agreement; and branches with $R^2 < 0.5$ were assigned Code~0.0, indicating poor agreement. The detailed flow diagram is summarized in Figure~\ref{fig:r2_combined_report}.

The four branch codes are combined in branch order to classify each \texttt{CT} experiment. The main classes are \texttt{NoEF} for no electroforming (0000), \texttt{NeEF} for negative voltage electroforming (0001), \texttt{PoEF} for positive voltage electroforming (0111), and \texttt{EF} for an already electroformed response (1111). Experiments containing Code~0.5 are labeled \texttt{UNCERTAIN}; all other combinations are grouped as \texttt{OTHER}, as summarized in Supplementary Table~\ref{tab:ef_classification}.

\phantomsection
\label{subsubsec:ivnl_features}
\noindent\textbf{Feature Group 2: I-V Non-linearity Features.}
For post-electroforming I-V non-linearity analysis in the second measurement block, the same $\sinh$ fitting procedure was applied to each branch in the low-voltage region ($|V| \leq 0.5$\,V), without a compliance current limit. Here, $a$ and $b$ are fitting parameters that scale the current and voltage \cite{Yang2008Memristive}, with $b$ serving as the key descriptor of low-voltage non-linearity. Further mathematical details of the I-V non-linearity validation are provided in Supplementary Section~\ref{supp-sec:Hysteresis features: I-V non-linearity}.

\phantomsection
\label{subsubsec:roff_ron_features}
\noindent\textbf{Feature Group 3: R$_{\text{off}}$/R$_{\text{on}}$ Features.}
For R$_{\text{off}}$/R$_{\text{on}}$ analysis, a frequently used read bias of 0.2\,V was used \cite{Stathopoulos2019}. The ArC ONE board records the actual applied voltage, which can deviate from the target 0.2\,V read bias; in some branches, no raw point falls within the selected read-voltage window. Therefore, for branches with reliable low-voltage fits ($R^2 \geq 0.95$), R$_{\text{off}}$/R$_{\text{on}}$ was evaluated from the fitted $\sinh$ curve at exactly \(\pm 0.2\)\,V. This applies a common read bias across branch pairs rather than relying on the nearest measured raw point. Further details of the fitting procedure and R$_{\text{off}}$/R$_{\text{on}}$ results are provided in Supplementary Sections~\ref{supp-sec:Roff/Ron features} and~\ref{supp-sec:Roff/Ron subset analysis}.

\phantomsection
\label{subsubsec:switching_features_main}
\noindent\textbf{Feature Group 4: Switching Dynamics Features.}
The programming segments from the third measurement block were used to quantify pulse-induced resistance changes under different voltage-stimulation conditions. For each switching segment of 200 pulses, the switching magnitude was calculated as the difference between the mean resistance of the first and the last 10 pulses. The switching surface data are included in the database to support the Verilog-A compact-model \cite{Messaris2018,Stathopoulos2019SwitchingUncertainty}. Further analysis is shown in Supplementary Section~\ref{supp-sec:switching_features} and~\ref{supp-sec:Switching magnitude summaries}. A graphical user interface (GUI) for generating switching surfaces \cite{Messaris2018} is provided, with details in \hyperref[sec:Usage Notes]{Usage Notes}.


\phantomsection
\label{subsubsec:volatility_features_main}
\noindent\textbf{Feature Group 5: Short-Term Volatility Features.}
For volatility analysis in the same third measurement block, each programming segment was followed by a non-invasive read segment to track post-programming resistance relaxation. For each volatility read segment, the volatility magnitude was calculated as the difference between the mean resistance of the first and the last 10 points within the read segment. The volatility and relaxation fitting results are included in the database to support the Verilog-A compact-model \cite{Giotis2020BidirectionalVolatile}. Further examples and volatility summaries are shown in Figure~\ref{fig:volatility_data} and Supplementary Section~\ref{supp-sec:Volatility magnitude summaries}.

\begin{figure}[h!]
    \centering
    \includegraphics[width=0.55\linewidth]{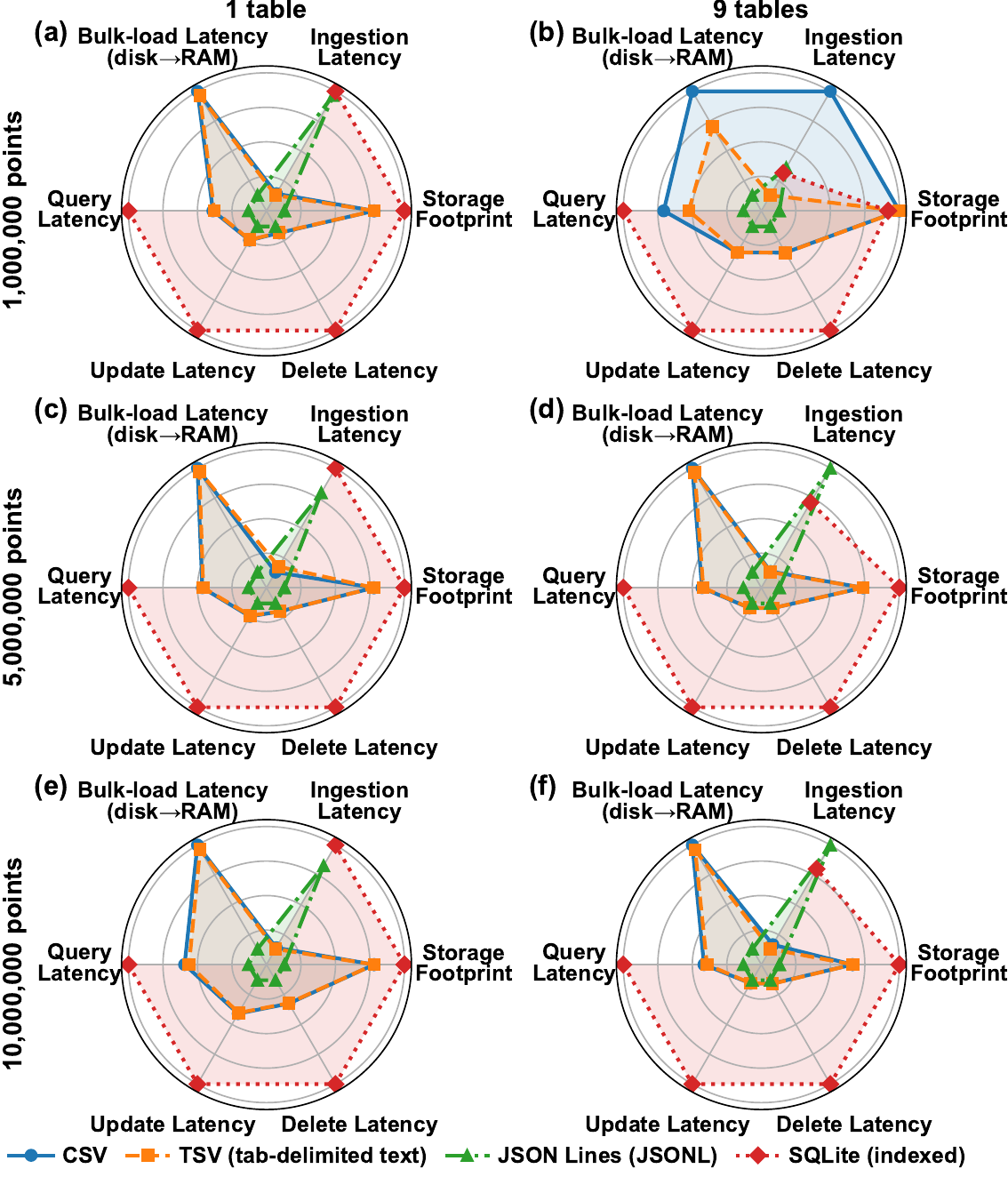}
    \caption{\textbf{Storage format and structure benchmarking.} Radar plots compare CSV, TSV, JSON Lines, and indexed SQLite across six performance metrics, with larger radii indicating better relative performance within each plot. Dataset sizes are (a,b) $10^6$, (c,d) $5\times10^6$, and (e,f) $10^7$ electrical point records. The left column shows a denormalized single-table structure, and the right column shows a normalized multi-table structure.}
    \label{fig:storage_benchmark}
\end{figure}

\phantomsection
\label{subsec:methods_storage_benchmarking}
\noindent\textbf{Storage benchmarking.}

The storage benchmark was designed to evaluate the need for a normalized relational database structure for large memristor datasets, rather than simply comparing file formats. In this dataset, electrical data point records are generated at a very large scale, while fabrication, wafer mapping, electrical characterization, and modeling must remain linked without being duplicated for every electrical data point. Figure~\ref{fig:storage_benchmark} summarizes the storage benchmarks performed on a synthetic source dataset. Here, data format refers to the storage format itself, namely CSV, TSV, JSON Lines, and indexed SQLite, whereas data structure refers to the organization of the content as either a fully denormalized single-table layout or a normalized multi-table hierarchy. In this context, normalization means storing shared metadata in separate linked tables rather than repeating them for every electrical data point \cite{Date2003,Codd1970}. Detailed benchmark definitions, hardware and software configuration, and full results are provided in Supplementary Section~\ref{subsec:Storage benchmarking}.

At a relatively large scale, the data format primarily determined the absolute performance level. Indexed SQLite was consistently the fastest format for queries, updates, and deletions, whereas JSON Lines became the slowest at larger sizes, especially for read-modify-rewrite workloads. CSV and TSV showed broadly similar behavior. At \(10^7\) electrical point records in the single-table benchmark, query latencies were 13.8\,s for CSV, 17.6\,s for TSV, 428.0\,s for JSON Lines, and 0.678\,s for SQLite; the corresponding multi-table values were 4.48\,s, 5.02\,s, 19.7\,s, and 0.216\,s. 

By contrast, data structure primarily affected storage footprint and scaling behavior. For the same source at \(10^7\) electrical point records, the normalized multi-table hierarchy reduced footprint from 2,065\,MiB to 762\,MiB for CSV/TSV, from 4,863\,MiB to 1,146\,MiB for JSON Lines, and from 1,543\,MiB to 588\,MiB for SQLite (1 MiB = 1024$^2$ bytes, 1 GiB = 1024$^3$ bytes). The same structural change also reduced query time by 3.09$\times$ for CSV, 3.51$\times$ for TSV, 21.8$\times$ for JSON Lines, and 3.13$\times$ for SQLite, consistent with normalization reducing repeated metadata and rewrite volume at larger scale \cite{Codd1970,ElmasriNavathe2016,Silberschatz2020}. 

The same structural requirement is reflected in the deployed database storage report (\hyperref[sec:data_availability]{Data Availability}). The normalized SQLite database occupies 13.85\,GiB, including 11.39\,GiB for tables and 2.46\,GiB for indices. Although the schema contains 37 tables and 336 columns, the \texttt{Experimental\_Detail} table contains only 8 electrical point columns yet already occupies 9.24\,GiB, with a further 2.04\,GiB from its main index linked to experiments. A fully denormalized structure that repeated broader metadata alongside each electrical point record would therefore expand the database to several hundred GiB. This confirms that the normalized hierarchy is not only a design preference, but a practical requirement for storing and reusing large memristor characterization datasets.

\section*{Data Records}

The deposited dataset is accompanied by supporting documents for inspection, interpretation, and reuse.

\begin{figure}[h!]
    \centering
    \includegraphics[width=0.6\linewidth]{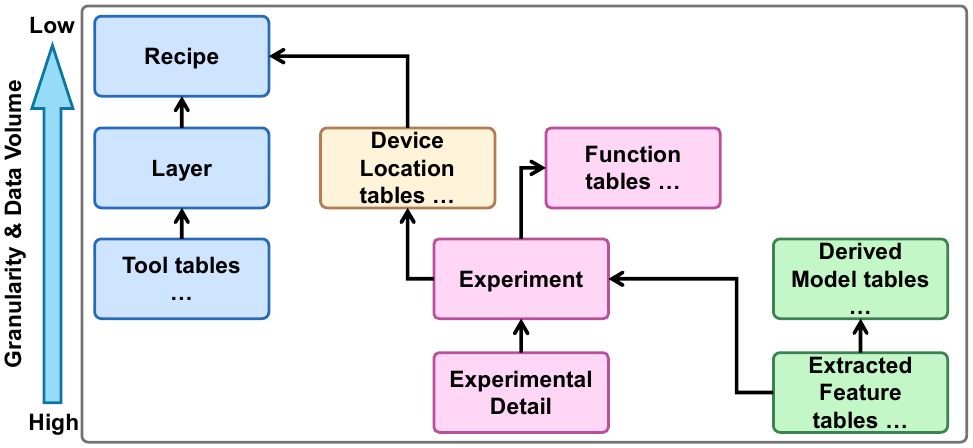}
    \caption{\textbf{Database structure and granularity.} Colors match the data-collection pipeline in Figure~\ref{fig:data_measure_procedure} (blue: Fabrication; yellow: Wafer Mapping; red: Characterization; green: Modeling). Vertical placement indicates data granularity and volume. Arrows denote foreign keys that link fabrication, locations, measurements, and features.}
    \label{fig:database_structure}
\end{figure}

Figure~\ref{fig:database_structure} provides a high-level overview of the table groups in the database structure. Consistent with the pipeline described in Figure~\ref{fig:data_measure_procedure}, the database is organized from lower-granularity contextual records, including fabrication, wafer mapping, electrical characterization, and modeling tables, to the high-volume \texttt{Experimental\_Detail} table, and then to the extracted feature tables and derived model tables that support analysis and reuse. This organization preserves traceable links from process context and physical device identity to raw electrical measurements and derived analytical results.

The name of the database is \texttt{	
Memristor\_Database.db}. Complete table- and column-level definitions are provided in the deposited repository through the accompanying database documentation files. These include \texttt{schema.sql} and \texttt{er\_diagram.svg} for the database structure, \texttt{table\_overview.csv} for summaries of the 37 database tables, \texttt{column\_dictionary.csv} for 336 column-level definitions, \texttt{core\_relationships.csv} for key foreign-key links and deletion rules, \texttt{selected\_qc\_rules.csv} for reader-facing integrity and quality-check rules, \texttt{unit\_suffix\_conventions.csv} for naming and unit-suffix conventions, \texttt{storage\_usage.txt} for the table- and index-level storage footprint of the deposited SQLite database, and a \texttt{README} file explaining the annotation package. Repository access and citation details are provided in the \hyperref[sec:data_availability]{Data Availability} section.

\section*{Technical Validation}

\textbf{Raw data validation.} Experimental records are filtered based on fabrication, wafer mapping, and electrical characterization integrity before being inserted into the normalized database. Of 192,176 initial experimental records, 161,006 were retained in the final curated dataset.

Within the fabrication workflow summarized in Figure~\ref{fig:data_measure_procedure}(a), experiments are excluded if: i) they cannot be linked to a user-defined fabrication assignment; or ii) the corresponding fabrication information was missing or insufficient in the development stage of the database. After this filtering, 189,106 experimental records were retained.


Within the wafer-mapping workflow summarized in Figure~\ref{fig:data_measure_procedure}(b), validation was used to confirm the device location mapping of each experimental record. Experiments were excluded if: i) they could not be linked to a device location; ii) the linked location could not be verified against the wafer map as a real die/sub-die/device position; or iii) the verified location fell outside the Stand-Alone (SA) layout considered in this dataset. After this filtering, 178,542 experimental records were retained.

Within the electrical characterization workflow summarized in Figure~\ref{fig:data_measure_procedure}(c), data validation was used to confirm that each experiment contained a usable electrical measurement record. Experimental records were excluded if: i) they could not be linked to a single measurement function in the metadata; ii) the associated electrical data points were missing, as can occur in real high-throughput automated measurements; iii) physically invalid resistance or voltage-amplitude values were present; or iv) the stored measurement metadata were inconsistent with the observed number of electrical data points for \texttt{CT}, \texttt{PF}, or \texttt{PF-IR}. After this filtering, the final curated dataset contained 161,006 experimental records.

Overall, the validation pipeline reduced the dataset from 192,176 candidate experiments to 161,006 retained experiments, but the associated \texttt{Experimental\_Detail} table decreased by only 1.09\% (from 171,134,064 to 169,270,462 rows). This shows that most exclusions were driven by metadata or record-completeness issues, rather than by the removal of large amounts of raw electrical data.

\begin{figure}[h!]
    \centering
    \makebox[\linewidth][c]{%
        \includegraphics[width=1.\linewidth]{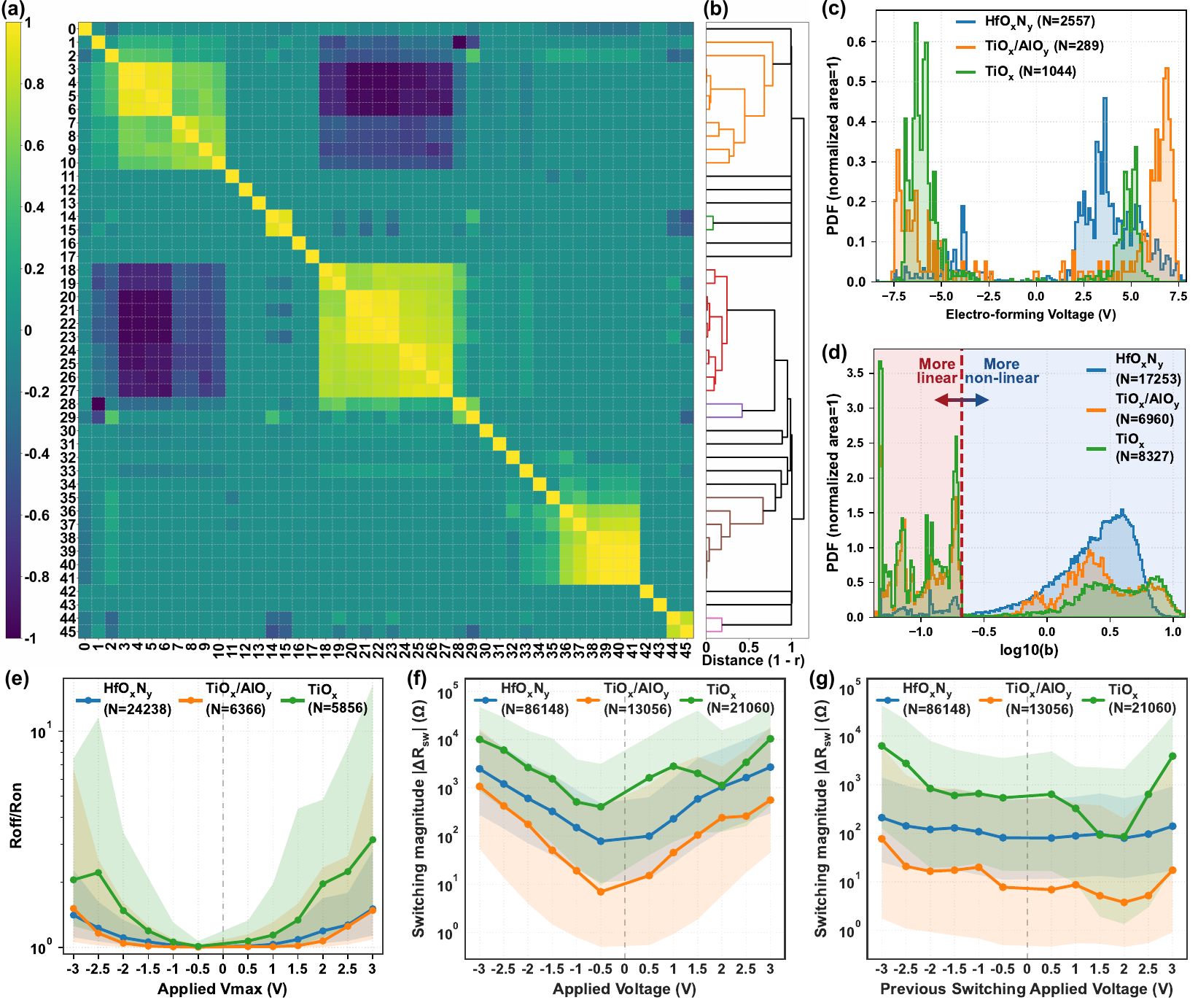}
    }
    \caption{\textbf{Feature correlations and distributions.} (a) Spearman cross-correlation heat map of aligned memristor features, ordered by hierarchical clustering. (b) Corresponding dendrogram. (c) Electroforming voltage distributions across recipe groups. (d) Distributions of fitted $\log_{10}(b)$ for electroformed memristors without compliance current hit. (e) The median of R$_{\text{off}}$/R$_{\text{on}}$ versus applied $V_\mathrm{max}$ for the higher-$b$ subset. (f) Absolute switching magnitude, $|\Delta R_{\mathrm{sw}}|$, versus applied voltage for the filtered \texttt{PF-IR} dataset. (g) Absolute volatility magnitude, $|\Delta R_{\mathrm{vol}}|$, versus preceding switching voltage for the same filtered \texttt{PF-IR} dataset.}
    \label{fig:cct_feature_relationships}
\end{figure}

\textbf{Feature data validation.} Feature-level validation examined whether the extracted features formed a coherent and interpretable feature space. Since memristor responses depend strongly on material stack, fabrication, and measurement protocol, prior reports were used as contextual support for expected trends rather than as one-to-one numerical benchmarks. Cross-feature correlations, hierarchical clustering, and representative recipe- and protocol-resolved trends were used to assess whether the feature tables captured complementary aspects of device response. The aim was not to identify a single mechanism or a single best-performing stack, but to confirm that the feature space provides a reliable basis for multidimensional comparison of device-response profiles across experimental contexts.

Figure~\ref{fig:cct_feature_relationships} summarizes the validation results for extracted features. Figure~\ref{fig:cct_feature_relationships}(a) shows a pairwise Spearman cross-correlation \cite{Spearman1904,HaukeKossowski2011} heatmap of the memristor features, with features ordered by hierarchical clustering. The mapping of numbered feature labels is provided in Supplementary Table~\ref{tab:feature_mapping}. Figure~\ref{fig:cct_feature_relationships}(b) shows the corresponding dendrogram, where hierarchical clustering groups features with similar variation patterns rather than predefined categories~\cite{MurtaghContreras2012}. Figures~\ref{fig:cct_feature_relationships}(a) and (b) demonstrate that the extracted features are not isolated metrics, but rather interconnected descriptors that form a cohesive multidimensional space. While this overarching correlation structure highlights the complex covariation in device behavior, we isolate the individual feature groups in Figures~\ref{fig:cct_feature_relationships}(c)-(g) to explicitly validate their specific distributions and protocol dependencies.

In Figure~\ref{fig:cct_feature_relationships}(c), electroforming voltage shows broad, recipe-dependent distributions rather than a single universal threshold, supporting its use as a recipe-sensitive feature and remaining consistent with context-dependent electroforming and filament formation, with details in Figures~\ref{fig:electroform_recipe_distribution}, \ref{fig:r_before_ohm_violin_by_recipe}, and~\ref{fig:ef_success_rate_vs_area_by_recipe}, and Supplementary Section~\ref{supp-sec:Electroforming distributions and grouping details} \cite{Waser2007Nanoionics,Kwon2010a,Nandi2020Electroforming}. 

In Figure~\ref{fig:cct_feature_relationships}(d), $\log_{10}(b)$ values from electroformed branches without compliance current hit show recipe-dependent low-voltage non-linearity; $b$ is used here as a curvature descriptor under the $\sinh$ fit in Equation\,\ref{eqn:sinh_fit_common}, with details in Figure~\ref{fig:iv_nonlinearity_fitting_distribution}, and Supplementary Section~\ref{supp-sec:Low-bias non-linearity summaries} \cite{Yang2008Memristive,Yakopcic2013,Messaris2018}. 

In Figure~\ref{fig:cct_feature_relationships}(e), the higher-$b$ subset shows clearer recipe and $|V_{\max}|$ dependence in R$_{\text{off}}$/R$_{\text{on}}$, whereas the lower-$b$ subset in Figure~\ref{fig:roff_ron_vmax_meanb_ab}(b) remains near unity. These trends reinforce the interpretation of R$_{\text{off}}$/R$_{\text{on}}$ as a memory window governed by both applied voltage and resistance state, consistent with TiO$_2$- and TiO$_2$/Al$_2$O$_3$-based studies where barrier design, defect profile, and programming conditions affect non-linearity, resistance states, and ON/OFF readout contrast \cite{Stathopoulos2019,Stathopoulos2017,Govoreanu2013}. Details are provided in Figures~\ref{fig:roff_ron_bsplit_combined_distribution} and~\ref{fig:roff_ron_vmax_meanb_ab}, and Supplementary Section~\ref{supp-sec:Roff/Ron subset analysis}.

In Figure~\ref{fig:cct_feature_relationships}(f) and Figure~\ref{fig:cct_feature_relationships}(g), switching and volatility magnitudes vary with the applied or preceding switching voltage, demonstrating that these dynamic descriptors remain highly sensitive to the programming stimuli. This is consistent with existing literature linking conductance updates, memory window evolution, and volatile dynamics to pulse conditions and stimulus history \cite{Frascaroli2018,Covi2021a,Moon2024a}. Details are provided in Figures~\ref{fig:switching_pdf_recipe_pubquality}, \ref{fig:switching_pdf_voltage_segment_pubquality_abs10_trial_narrow_n}, \ref{fig:volatility_pdf_recipe_pubquality_excluding_allseg_abs_sw_lt10}, \ref{fig:volatility_pdf_voltage_segment_pubquality_abs10}, and~\ref{fig:vol_delta_abs_over_sw_delta_abs_vs_rstart_binned20_median_by_recipe_pubquality_abs10}, and in Supplementary Sections~\ref{supp-sec:Switching magnitude summaries} and~\ref{supp-sec:Volatility magnitude summaries}.

\section*{Usage Notes}
\label{sec:Usage Notes}

The deposited SQLite database is the primary reusable record, accompanied by repository documentation defining the tables, fields, units, and relationships.

\textbf{For direct reuse}, the database supports SQL/Python querying, joining, aggregation, and custom analysis across fabrication, wafer mapping, electrical characterization, and modeling tables.

The graphical interfaces shown below provide optional examples of database reuse, including record inspection, visualization, diagnostics, and subset export.

\begin{figure}[h!]
    \centering
    \includegraphics[width=0.94\linewidth]{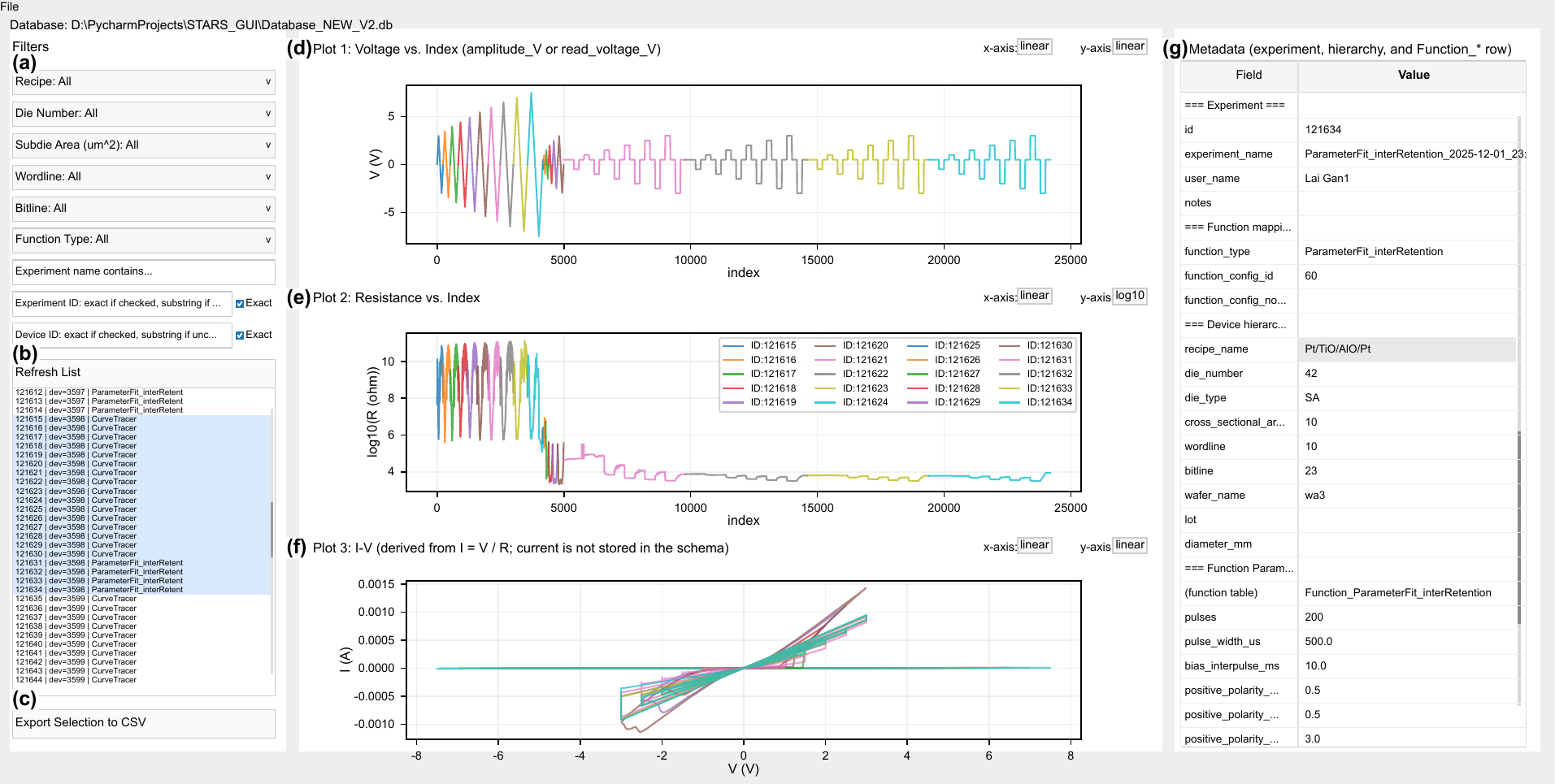}
    \caption{\textbf{Raw-data handling interface and visualizer.} The interface supports inspection and export of experimental records through (a) filters, (b) experiment selection, (c) CSV export, (d)-(f) raw-data visualization, and (g) linked metadata inspection.}
    \label{fig:raw_data_browser_gui}
\end{figure}

\textbf{Raw-data browser GUI.} Figure~\ref{fig:raw_data_browser_gui} shows an integrated interface for querying, selecting, exporting, and visualizing experimental measurement records. Figure~\ref{fig:raw_data_browser_gui}(a) contains filter controls for recipe, die number, sub-die area, wordline, bitline, function type, experiment name, experiment ID, and device ID. Figure~\ref{fig:raw_data_browser_gui}(b) lists the matching experiments and supports single- or multi-record selection. Figure~\ref{fig:raw_data_browser_gui}(c) exports the selected experiments to CSV together with associated database records. Figures~\ref{fig:raw_data_browser_gui}(d)-(f) plot the voltage, resistance, and derived current-voltage (I-V) characteristics of the selected record. In the I-V panel, the current is calculated as $I = V/R$. Figure~\ref{fig:raw_data_browser_gui}(g) summarizes linked metadata from the experiment record, fabrication, device location, and the associated function parameter record. In the presented implementation, database access is read-only, function metadata and linked context are preloaded once and then filtered in computer memory, whereas electrical points are fetched only for the selected experimental records. This interface therefore provides a practical route for interactive inspection, tracing, visualization, and lightweight subset export.

\begin{figure}[h!]
    \centering
    \includegraphics[width=0.94\linewidth]{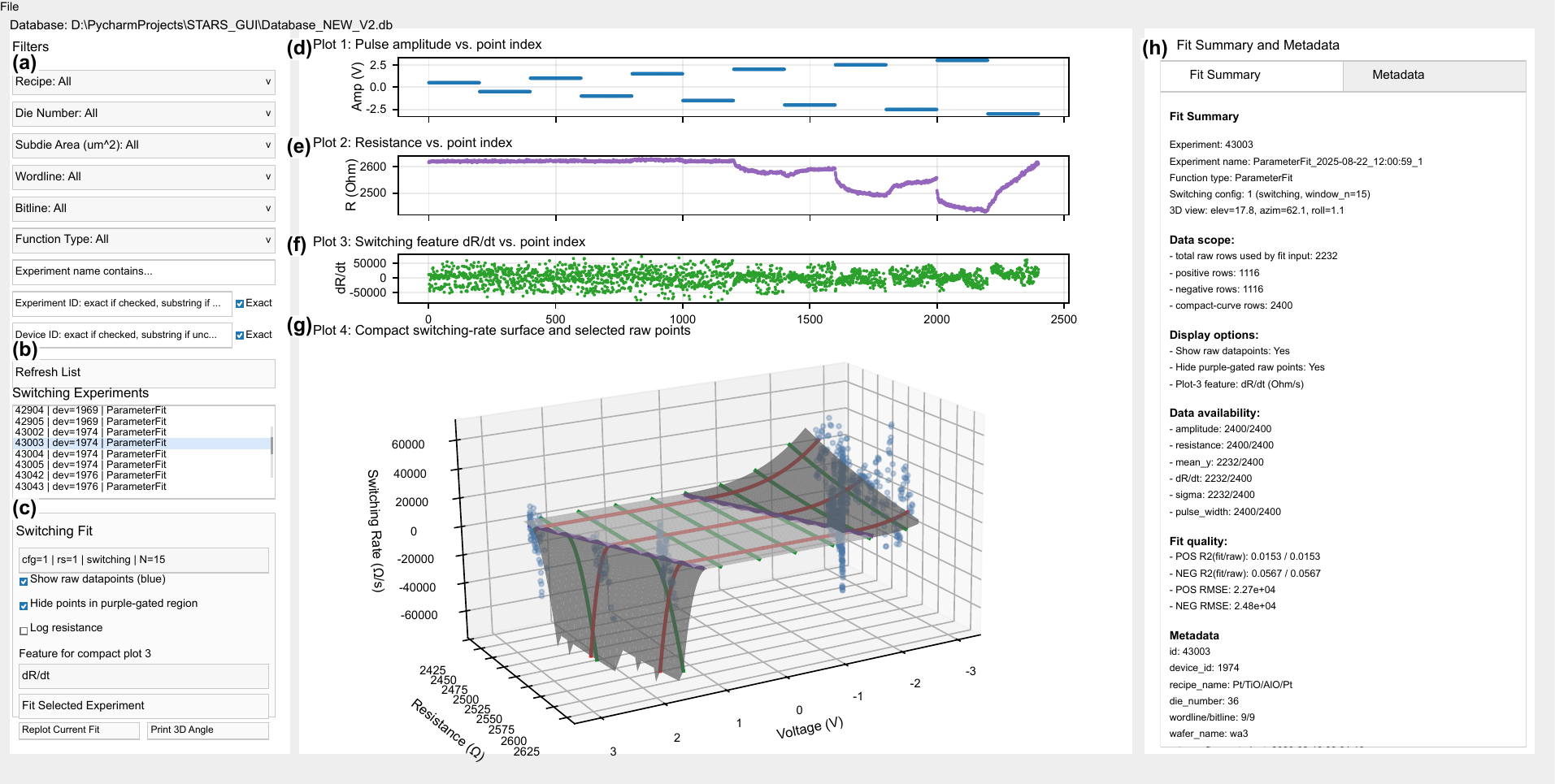}
    \caption{\textbf{Resistance-switching feature fitting interface.} The interface supports (a) search and (b) selection of switching experiments, (c) fitting setup and display control, (d)-(f) inspection of diagnostic pulse-sequence traces, (g) voltage-resistance-$dR/dt$ surface fitting, and (h) review of fit quality and linked metadata.}
    \label{fig:switching_fit_gui}
\end{figure}

\textbf{Feature and modeling GUI.} Figure~\ref{fig:switching_fit_gui} shows an interface for filtering switching experiments with linked switching rate results, fitting positive and negative switching segments, and reviewing the resulting diagnostics together with metadata information. The fitting workflow follows the switching rate modeling approach used to describe voltage- and resistance-dependent changes in memristor devices and can be implemented in Verilog-A or other compact modeling systems~\cite{Messaris2018,Stathopoulos2019SwitchingUncertainty}. Figures~\ref{fig:switching_fit_gui}(a) and (b) are used to filter the database and select a target switching experimental record. Figure~\ref{fig:switching_fit_gui}(c) defines the fitting configuration and display options, including switching-rate configuration selection, feature display, raw-point visibility, resistance scaling, replotting, view-angle reporting, and PNG saving. Figures~\ref{fig:switching_fit_gui}(d)-(f) provide diagnostic pulse-sequence traces against ordered pulse-point index, showing pulse amplitude, resistance, and a selected switching feature. Figure~\ref{fig:switching_fit_gui}(g) presents the switching behavior mapped across three dimensions: voltage, resistance, and the switching rate ($dR/dt$). In this view, raw data points are overlaid on the generated switching surfaces to enable direct visual comparison. Figure~\ref{fig:switching_fit_gui}(h) reports fit summary and linked metadata, including data scope, data availability, fit-quality metrics, experiment information, function parameters, and switching rate configuration. Together, these views demonstrate how deposited records can be integrated into downstream model fitting and validation workflows, all while preserving strict traceability to their original measurement context. In its current implementation, this interface is designed for exploratory fitting, diagnostics, and data review, rather than serving as a static, pre-computed model layer like the extracted feature tables.

Further implementation details for these graphical interfaces are provided with the released code and accompanying documentation described in the \hyperref[sec:code_availability]{Code Availability} section.

\section*{Data Availability}
\phantomsection
\label{sec:data_availability}

The dataset supporting this study has been deposited in Zenodo under the
reserved DOI \href{https://doi.org/10.5281/zenodo.19390197}{10.5281/zenodo.19390197}.
The Zenodo record is currently under restricted access and will be made publicly
available before publication of the peer-reviewed journal article.

\section*{Code Availability}
\label{sec:code_availability}

The code used for data processing, validation, and visualization is available at \url{https://github.com/aprilaihub/STARS_GUI}.

\section*{Author Contributions}
L.G., B.D.R., and T.P. conceived the study. L.G., G.H., D.Y., and B.D.R. contributed to data collection, curation, discussion of content, writing, and editing of the manuscript. S.S., B.D.R., and T.P. supervised the work and contributed to reviewing and editing the manuscript before submission. All authors reviewed and approved the final manuscript.

\section*{Competing Interests}
The authors declare no competing interests.

\section*{Acknowledgements}
The authors wish to thank the students and staff at the Centre for Electronics Frontiers and the Scottish Microelectronics Centre for their invaluable technical contributions. In particular, Dr Alin Panca, Dr Hannah Levene, and Dr Peter Lomax. 

\section*{Funding}
This work was supported by UK Research and Innovation (UKRI) and the Engineering and Physical Sciences Research Council (EPSRC) through the AI Hub for Productive Research and Innovation in Electronics (APRIL AI Hub), grant number EP/Y029763/1.


\clearpage

\setcounter{section}{0}
\setcounter{figure}{0}
\setcounter{table}{0}
\setcounter{equation}{0}

\renewcommand{\thesection}{S\arabic{section}}
\renewcommand{\thesubsection}{S\arabic{section}.\arabic{subsection}}
\renewcommand{\thesubsubsection}{S\arabic{section}.\arabic{subsection}.\arabic{subsubsection}}
\renewcommand{\thefigure}{S\arabic{figure}}
\renewcommand{\thetable}{S\arabic{table}}
\renewcommand{\theequation}{S\arabic{equation}}

\renewcommand{\theHsection}{supp.\arabic{section}}
\renewcommand{\theHsubsection}{supp.\arabic{section}.\arabic{subsection}}
\renewcommand{\theHsubsubsection}{supp.\arabic{section}.\arabic{subsection}.\arabic{subsubsection}}
\renewcommand{\theHfigure}{supp.\arabic{figure}}
\renewcommand{\theHtable}{supp.\arabic{table}}
\renewcommand{\theHequation}{supp.\arabic{equation}}

\begin{center}
{\LARGE\bfseries Supplementary Information\par}
\vspace{0.6em}
{\large A relational fabrication-to-modeling database for memristor devices\par}
\end{center}

\tableofcontents
\newpage

\section{Fabrication}
\subsection{Fabrication details}
\label{supp-sec:fabrication}

\begin{table}[!h]
\centering
\caption{Fabrication parameters for BE/TE and Insulator Layer}
\label{tab:fabrication_parameters}

\small
\setlength{\tabcolsep}{2pt}
\renewcommand{\arraystretch}{1.25}

\begin{tabular}{|c|c|c|c|>{\centering\arraybackslash}p{4cm}|c|c|c|>{\centering\arraybackslash}p{4cm}|}
\hline
\textbf{Device} &
\multicolumn{4}{c|}{\textbf{BE/TE}} &
\multicolumn{4}{c|}{\textbf{Insulator Layer}} \\
\hline

& \textbf{Mat} & \textbf{Thick} &
\textbf{Tool} & \textbf{Deposition Details} &
\textbf{Mat} & \textbf{Thick} &
\textbf{Tool} & \textbf{Deposition Details} \\

&  & \textbf{(nm)} &  &
&  & \textbf{(nm)} &  & \\
\hline

\textbf{D1}
& TiN & 50/50 & SPT
& Target: Ti; Pressure: 2 mT; Gun: DC, Ar: 20 sccm; N: 20 sccm. 
& HfO$_\text{x}$N$_\text{y}$ & 10 & ALD
& Precursors: TDMAHf / H$_2$O / N$_2$ = 1:1:1; Method: thermal; Deposition temp: 250$^\circ$C \\
\hline

\textbf{D2}
& Pt &  12/20 & EB
& Target: Pt; Pressure: 3 mT; Deposition rate: 0.06 nm/s
& TiO$_\text{x}$/AlO$_\text{y}$ & 24/4 & SPT
&  TiO$_\text{x}$ - Target: Ti; Pressure: 3 mT; Gun: DC; Ar: 90 sccm; O: 6.6 sccm; Power density: 2.5 $W/cm^2$

AlO$_\text{y}$ - Target: Al; Pressure: 1 mT; Gun: PDC; Frequency: 30 kHz; Ar: 50 sccm; O: 12 sccm; Power density: 3.2 $W/cm^2$ \\
\hline

\textbf{D3}
& Pt &  12/20 & EB
& Target: Pt; Pressure: 3 mT; Deposition rate: 0.06 nm/s
& TiO$_\text{x}$ & 25 & SPT
& Target: Ti; Pressure: 8 mT; Gun: DC; Ar: 20 sccm; O: 1 sccm; Power density: 3.2 $W/cm^2$ \\
\hline

\end{tabular}

\vspace{2pt}
\begin{minipage}{\linewidth}
\footnotesize
\raggedright
\textit{Note:} BE and TE denote bottom and top electrode, respectively; both electrodes share identical fabrication parameters and are therefore grouped under the same column. Mat refers to material and Thick refers to thickness. ALD denotes atomic layer deposition, EB denotes E-beam evaporation, and SPT denotes sputtering. TDMAHf denotes tetrakis(dimethylamino)hafnium. PDC denotes pulsed direct current. 
\end{minipage}

\normalsize
\end{table}

Table\,\ref{tab:fabrication_parameters} summarizes material stack information for the three device technologies used in this work. The detailed information is stored in the main database and released on Zenodo.

For each fabrication step, the original vendor recipe file (e.g., .txt, .seq, or .rcp) is retained verbatim in the central database as a Binary Large Object (BLOB) \cite{SQLiteDatatype,SQLiteBlobIO}. The BLOB provides an immutable provenance record, but the binary or semi-structured file content is not always suitable for direct querying, comparison, or analysis, and may not fully reflect the final device properties across the wafer. Each BLOB is therefore linked via a foreign key \cite{SQLiteForeignKeys} to a curated recipe record that stores structured recipe parameters checked through the wafer-creation workflow and repeated verification with the fabrication engineers. This design separates preservation of the original vendor file from the structured, validated recipe metadata used for downstream retrieval and analysis.

Within each curated recipe record, information is captured in two forms. First, the database stores structured process metadata, including materials, process gases, thicknesses, flow rates, temperatures, and other relevant process parameters. Second, for ALD recipes, the process sequence is stored as an ordered set of nodes, following an adjacency-list model \cite{Celko2012Trees}. Each node represents a cycle, material, or gas step and records its parent node and position in the list. Detailed properties of cycles, materials, and gases are stored separately. This structure keeps the recipe compact, readable, and avoids duplicating the contents of repeated loops.

For inspection and editing, the curated recipe records described above are accessed through a lightweight GUI. These records include both structured process metadata and ALD sequence information. The recipe editor provides a rapid view of the stack configuration and tool arrangement, as shown in Figure~\ref{fig:recipe_construction_gui}. The \texttt{Link...} button controls the file-attachment interface for managing associations between recipe records and original vendor files, as shown in Figure~\ref{supp:file_attachment_interface}. The \texttt{Inserting material} control opens the ALD super-cycle editor, which provides an intuitive view of the nested super-cycle structure, as shown in Figure~\ref{supp:ald_cycle_editor}. These interfaces are provided as example access tools for inspecting and editing the database records, rather than as required components to make use of the database schema.
\newpage

\subsection{Fabrication GUI}
\label{supp-sec:Fabrication GUI}

\begin{figure}[h!]
    \centering
    \includegraphics[width=0.78\linewidth]{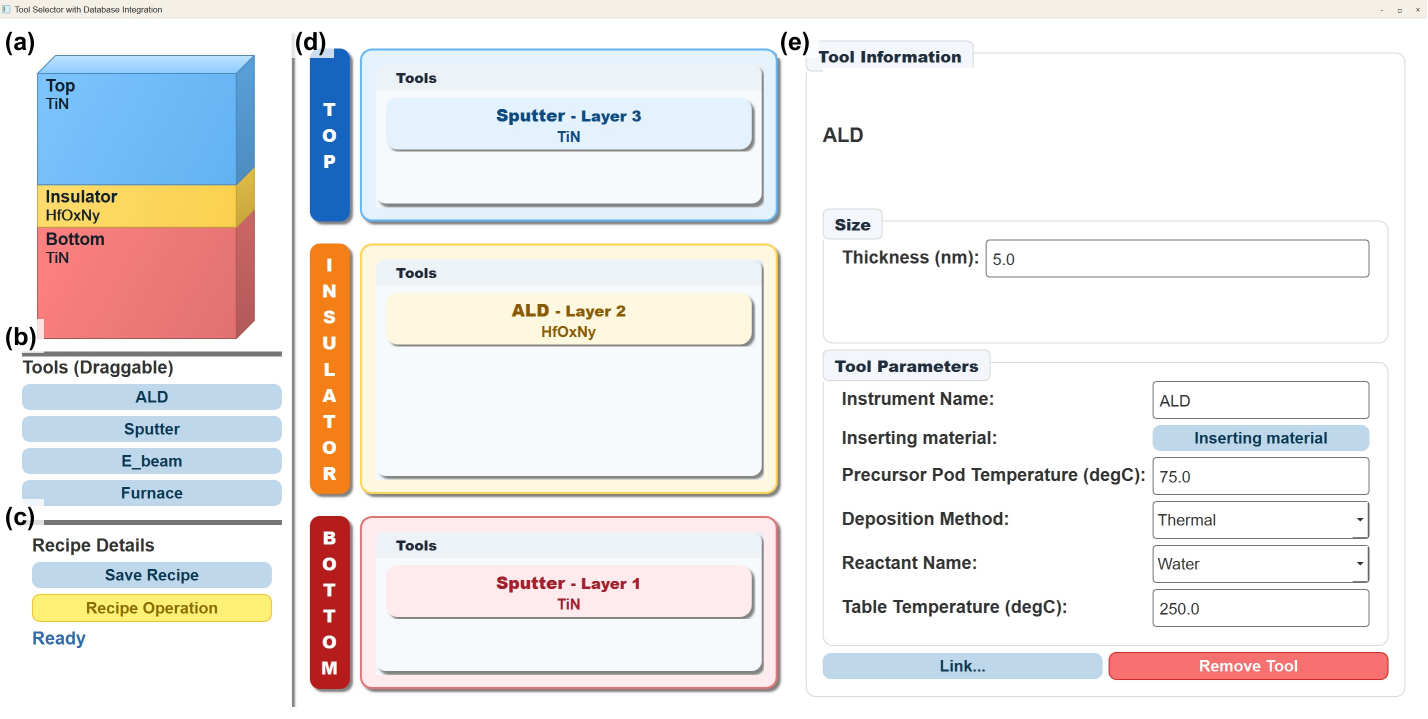}
    \caption{\textbf{Database-integrated material-layer recipe construction interface.} The interface supports fabrication recipe definition through five sections: (a) material and stack preview, (b) draggable tool list, (c) recipe operations, (d) layers of the tools, and (e) selected tool information and editable process parameters.}
    \label{fig:recipe_construction_gui}
\end{figure}

\textbf{Recipe-construction GUI.} Figure~\ref{fig:recipe_construction_gui} shows an interface for assembling, editing, and saving organized fabrication recipes. Figure~\ref{fig:recipe_construction_gui}(a) visualizes the current Top/Insulator/Bottom material stack as a structured preview of the assigned process steps. Figure~\ref{fig:recipe_construction_gui}(b) lists available fabrication tools, including ALD, sputtering, e-beam evaporation, and furnace/annealing processing, which can be inserted into the recipe workflow. Figure~\ref{fig:recipe_construction_gui}(c) provides recipe operations for saving the current working recipe and opening load, replace, or delete actions for stored recipes. Figure~\ref{fig:recipe_construction_gui}(d) organizes the process workflow by Top, Insulator, and Bottom layers, allowing individual fabrication steps to be assigned to specific layers. Figure~\ref{fig:recipe_construction_gui}(e) displays the selected tool information and editable process parameters, including thickness, material- or gas-related entries, and linked recipe file or ALD cycle editing. In the example shown in Figure~\ref{fig:recipe_construction_gui}, an ALD step is assigned to the HfO$_\text{x}$N$_\text{y}$ insulator layer, illustrating how the fabrication parameters for the three material stacks (TiN/HfO$_\text{x}$N$_\text{y}$/TiN, Pt/TiO$_\text{x}$/AlO$_\text{y}$/Pt, and Pt/TiO$_\text{x}$/Pt), can be captured in a structured form and linked to provenance database records. In the present implementation, this interface is intended for constructing and editing fabrication process flows.

\newpage

\begin{figure}[h!]
    \centering
    \includegraphics[width=0.55\linewidth]{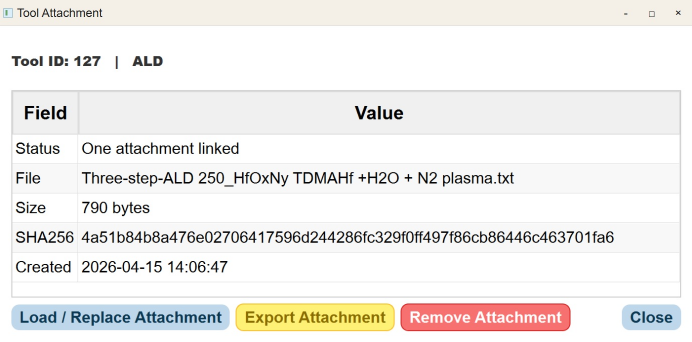}
    \caption{\textbf{Recipe-file attachment interface.} The interface is opened through the \texttt{Link...} button in the tool information section of Figure~\ref{fig:recipe_construction_gui}(e). It displays the linked recipe file, file size, SHA256 checksum, and creation time, and supports loading, replacing, exporting, and removing tool-level recipe-file attachments.}
    \label{supp:file_attachment_interface}
\end{figure}

\textbf{Recipe-file attachment interface.} Figure~\ref{supp:file_attachment_interface} shows the recipe file attachment interface opened from the \texttt{Link...} button in Figure~\ref{fig:recipe_construction_gui}(e). This interface records the recipe file associated with the selected fabrication tool and displays basic information, including the linked file name, file size, SHA256 checksum, and creation time. When a file is loaded, the checksum is used to verify whether an identical file has already been stored in the database; if a match is found, the existing file record can be linked rather than saving a duplicate copy. The interface also supports replacing an attachment, exporting the stored file, and removing an existing association. This provides a practical way to preserve vendor- or tool-specific recipe files alongside the structured process metadata stored in the database while reducing redundant file storage.

\newpage

\begin{figure}[h!]
    \centering
    \includegraphics[width=0.78\linewidth]{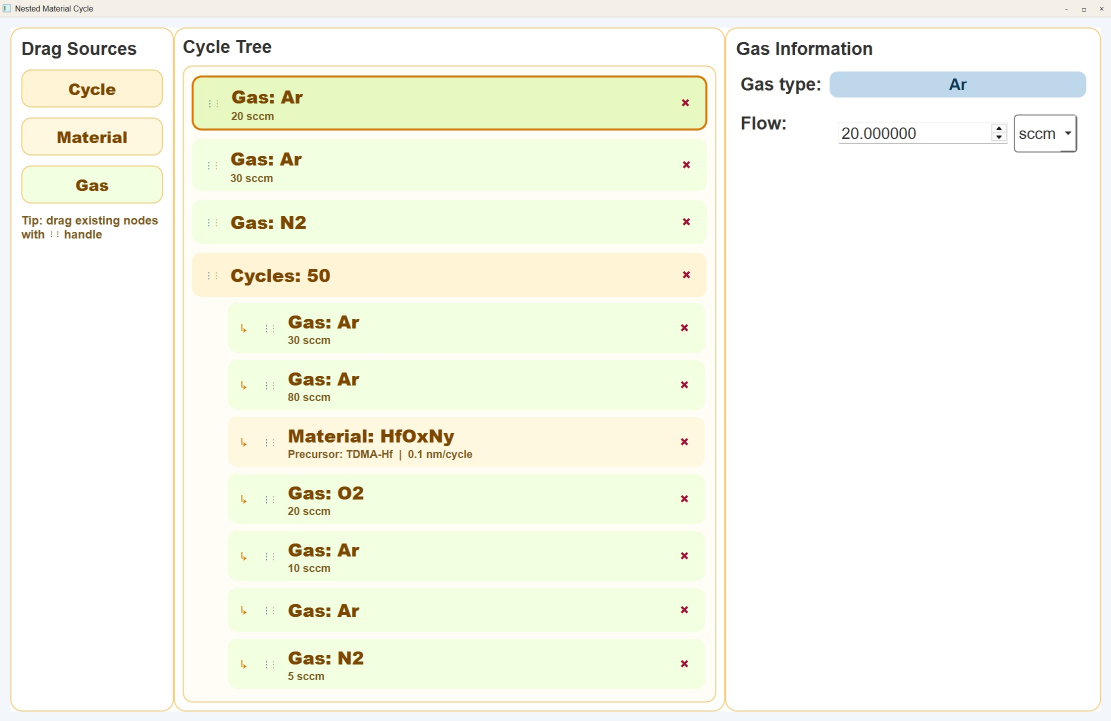}
    \caption{\textbf{ALD cycle editor for nested material-cycle structures.} The editor is opened through the \texttt{Inserting material} button in the tool information section of Figure~\ref{fig:recipe_construction_gui}(e). The interface provides draggable cycle, material, and gas nodes, a nested cycle tree for super-cycle construction, and editable parameters for the selected node.}
    \label{supp:ald_cycle_editor}
\end{figure}

\textbf{ALD cycle editor.} Figure~\ref{supp:ald_cycle_editor} shows the ALD cycle editor opened from the \texttt{Inserting material} button in Figure~\ref{fig:recipe_construction_gui}(e). The editor provides draggable cycle, material, and gas nodes, a central cycle tree, and a parameter panel for the selected node. This layout allows ALD super-cycles to be inspected and edited as nested parent-child structures. The interface provides a visual editing layer for the ALD structural records while preserving the same database used for recipe storage.

\newpage

\section{Wafer heatmap}
\begin{figure}[!htbp]
    \centering
    \includegraphics[width=1\linewidth]{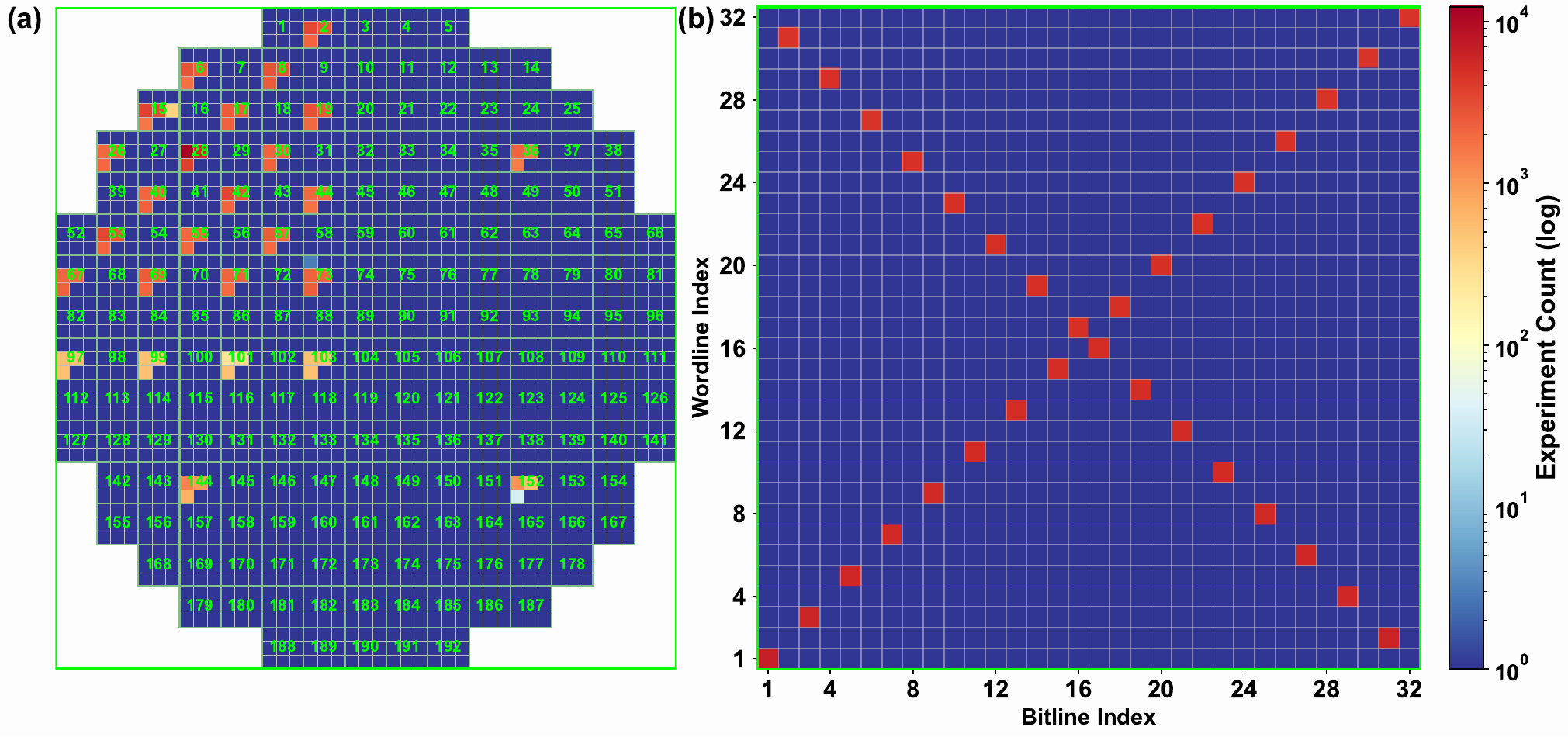}
    \caption{
    Experimental-record coverage at the wafer and device-array levels.
    (a) Wafer-level heatmap. (b) wordline-bitline device heatmap.
    }
    \label{fig:wafer_heatmap}
\end{figure}

Figure~\ref{fig:wafer_heatmap} shows the distribution of experimental records across the database at two spatial scales. In Figure~\ref{fig:wafer_heatmap}(a), the full wafer layout is retained, with each die divided into a \(3 \times 3\) sub-die grid corresponding to nine distinct device cross-sectional areas (CSAs). In Figure~\ref{fig:wafer_heatmap}(b), records are pooled across all recipes, dies, and sub-dies and mapped onto the wordline-bitline coordinates of the device array. Both panels use the same logarithmic color scale, allowing the wafer-level and device-level record densities to be compared directly.
\newpage

\section{Measurement protocol}
\label{supp-sec:Measurement protocol}
The standard measurement set was designed to extract five feature groups: electroforming, low-bias I-V non-linearity, Roff/Ron, switching dynamics, and short-term volatility. These groups were chosen to reflect the main stages of memristor characterization: initial activation, post-forming readout, pulse-driven switching, and post-pulse relaxation.

\begin{enumerate}
    \item \textbf{Electroforming} features are included since many oxide-based memristors require an initial activation step associated with the formation or stabilization of conductive pathways, and this transition can influence later switching behavior and variability \cite{Waser2007Nanoionics,Nandi2020Electroforming}.
    \item \textbf{I-V non-linearity} features are included since post-forming low-bias transport is a basic electrical descriptor of device state \cite{Qin2023a,Lanza2019RecommendedMethods}.
    \item \textbf{R$_{\text{off}}$/R$_{\text{on}}$} is included since it quantifies the separation between the programmed OFF- and ON-state resistances under read conditions, which is relevant to state distinguishability during readout and, in array-oriented contexts, to read-margin considerations \cite{Qin2023a,Wang2019a,Kang2022ClusterType,Gao2016SelfRectifying}.
    \item \textbf{Switching dynamics} are included since pulse-based measurements reveal resistance-update kinetics under programming conditions that cannot be fully inferred from quasi-static voltage sweeps alone \cite{Covi2021a}.
    \item \textbf{Volatility} features are included since post-pulse relaxation distinguishes stable non-volatile behavior from metastable or dynamic responses, which matters both for retention-oriented assessment and for applications that exploit short-term temporal behavior \cite{Covi2021a,Moon2024a}.
\end{enumerate}


These feature groups are derived from three sequential measurement blocks. The first block targets electroforming, the second targets post-forming low-voltage I-V response and R$_{\text{off}}$/R$_{\text{on}}$, and the third targets pulse-driven resistance evolution and post-pulse relaxation.
\newpage

\subsection{Measurement block one: electroforming with CurveTracer}
\label{supp-sec:Measurement block one}

The first two measurement blocks were performed using the standard \texttt{CurveTracer} (\texttt{CT}) function in the ArC ONE module \cite{ArcOneProduct,ArcOneUserGuide}, which implements a triangular-pulsed I-V measurement and supports an optional compliance-current limit. The \texttt{CT} function uses a programmable voltage step, start voltage, sweep limits, pulse width, bias type, and an optional compliance-current limit. In both blocks, the fixed \texttt{CT} settings were \texttt{voltage\_step\_V}=0.05\,V, \texttt{start\_voltage\_V}=0.05\,V, \texttt{step\_width\_ms}=50, \texttt{bias\_type}=Staircase, and \texttt{iv\_span}=Start towards V+.

The first block comprises 10 electroforming-oriented \texttt{CT} pulse trains. A symmetric compliance-current limit of 10\,\si{\micro\ampere} is applied to protect the device during the electroforming process. The only parameter varied between experiments is the symmetric positive/negative maximum voltage, which is increased from 3.0 to 7.5\,V in 0.5\,V steps. The applied-voltage pulse sequence is illustrated in Figure~\ref{fig:ef_data}(a).

\subsection{Measurement block two: I-V non-linearity and R$_{\text{off}}$/R$_{\text{on}}$ with CurveTracer}
\label{supp-sec:Measurement block two}
The second block comprises 6 post-electroforming \texttt{CT} pulse trains, with compliance current disabled. The compliance current limitation was removed in this block to capture the post-forming I-V hysteresis response unconstrained by current compliance, from which I-V non-linearity and R$_{\text{off}}$/R$_{\text{on}}$ were extracted. The only parameter varied between experiments is the symmetric positive/negative maximum voltage, which is increased from 0.5 to 3.0\,V in 0.5\,V steps.  The corresponding applied-voltage sequence is shown in Figure~\ref{fig:hys_data}(a).

\subsection{Measurement block three: switching and volatility with ParameterFit and ParameterFit\_interRetention}
\label{supp-sec:Measurement block three}
The third measurement block comprised 4 pulse trains for assessing switching and volatility, executed using the \texttt{ParameterFit} (\texttt{PF}) and \texttt{ParameterFit\_interRetention} (\texttt{PF-IR}) functions. Across these experiments, the only parameter varied was the pulse width, set to 10, 50, 100, and 500\,\si{\micro\second}.

For experiments executed with \texttt{PF}, the fixed settings include \texttt{pulses}=200, \texttt{bias\_interpulse\_ms}=20, and symmetric positive/negative pulse-amplitude ranges from $\pm 0.5$ to $\pm 3.0$\,V in 0.5\,V steps. 

For experiments executed with \texttt{PF-IR}, the same alternating programming-voltage sequence is used; however, each programming segment is followed by a non-invasive read segment. The fixed settings include \texttt{pulses}=200, \texttt{bias\_interpulse\_ms}=10, the same symmetric positive/negative pulse-amplitude ranges, and read settings of \texttt{sr\_vread}=0.5\,V, \texttt{sr\_gap\_ms}=10, and \texttt{sr\_count}=200. This read-interleaving allows both switching evolution and post-pulse resistance stability to be captured within a single pulse train. The corresponding pulse sequences for \texttt{PF} and \texttt{PF-IR} are shown in Figures~\ref{fig:pf_data}(a) and \ref{fig:volatility_data}(a), respectively.
\newpage

\section{Memristor features from Electrical Characterization}
The following sections describe how the five feature groups are extracted from the raw measurement data. These features are used to compare device responses under defined measurement protocols, rather than to claim a unique physical mechanism for each response.
\subsection{Electroforming features}
\begin{figure}[!htbp]
    \centering
    \includegraphics[width=0.6\linewidth]{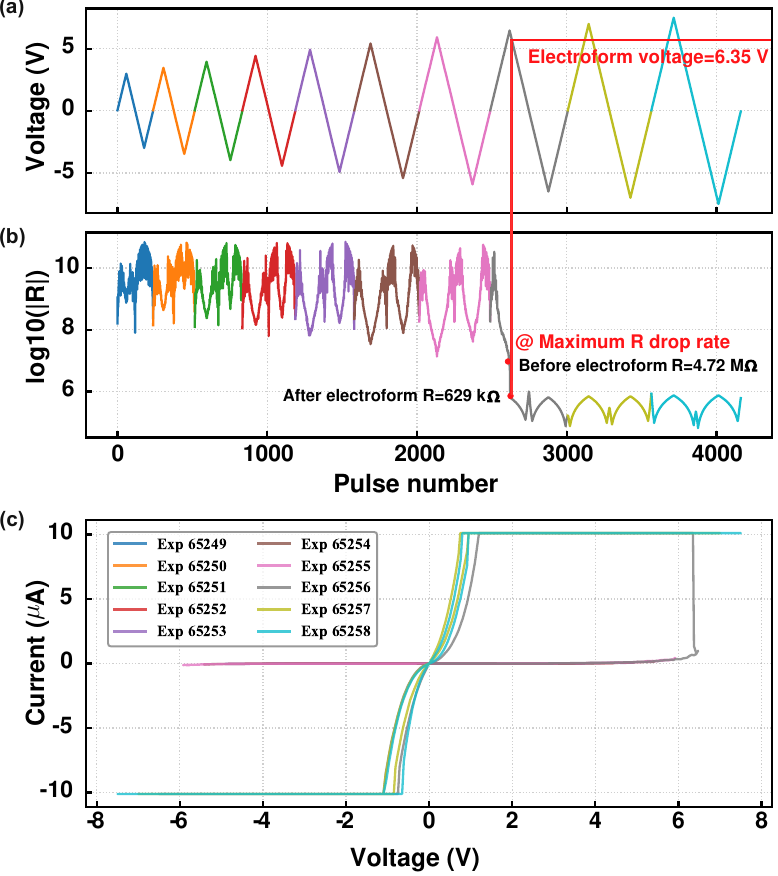}
    \caption{Example electroforming (EF) pulse train and measurement data. Trace colors are consistent across the three plots. (a) Applied voltage pulse train. Pulses are 100\,$\mu$s in duration with 10\,ms intervals. (b) Raw resistance data against applied voltage, plotted with log10 scale on y-axis. (c) EF curves in I-V plane. Example electroforming (EF) features for memristor. The four features are resistive state before EF, resistive state after EF, voltage at EF event, and drop ratio at EF event.}
    \label{fig:ef_data}
\end{figure}

Electroforming features provide a compact summary of the transition from an initially high-resistance, unformed device response to a more conductive state during electroforming. As electroforming is widely recognized to be highly variable and strongly device-dependent, fixed-threshold descriptions alone are often insufficient to capture its behavior across a heterogeneous dataset \cite{Lanza2019RecommendedMethods,Kwon2010a,Skaja2018}. Figure~\ref{fig:ef_data} illustrates the extraction procedure.

 In Figure~\ref{fig:ef_data}(a), the applied voltage pulse train is shown, with pulse amplitude increasing stepwise during the experiment. Figure~\ref{fig:ef_data}(b) shows the corresponding raw resistance evolution on a $\log_{10}$ scale, highlighting the abrupt resistance change associated with the electroforming event. Figure~\ref{fig:ef_data}(c) shows the corresponding I-V curves for the same experiment sequence. From these measurements, four electroforming descriptors were extracted: the resistive state before electroforming, the resistive state after electroforming, the electroforming voltage, and the resistance-drop ratio at the electroforming event.

For experiments assigned to the electroforming classes, negative voltage electroforming (\texttt{NeEF}) and positive voltage electroforming (\texttt{PoEF}), the electroforming voltage was determined from the raw measured voltage points. A compliance-current hit was defined as at least two consecutive raw points reaching the compliance threshold. When such a hit was present, the electroforming voltage was taken as the voltage at the first raw point reaching the compliance threshold. Otherwise, the electroforming voltage was assigned as the voltage at which the largest resistance drop occurred, as illustrated in Figure~\ref{fig:ef_data}(b). The resistive states before and after electroforming were taken from the resistance values immediately before and after the identified electroforming event. The drop ratio was calculated from the ratio of resistive state before and after the electroforming event. In this way, the extracted quantities remain tied to the raw experimental sequence rather than to an overly idealized or manually selected transition point.

These electroforming features summarize the observed transition in the recorded voltage-sweep sequence. A higher electroforming voltage indicates that a larger applied bias was required before the transition was detected, while a larger resistance-drop ratio indicates a greater change in conductance. The values can be affected by measurement noise, voltage-step resolution, compliance-current effects, and gradual rather than abrupt transitions, and are therefore used as descriptive summaries for large-scale comparison.
\newpage

\subsection{Hysteresis features: I-V non-linearity}
\label{supp-sec:Hysteresis features: I-V non-linearity}
\begin{figure}[!htbp]
    \centering
    \includegraphics[width=0.8\linewidth]{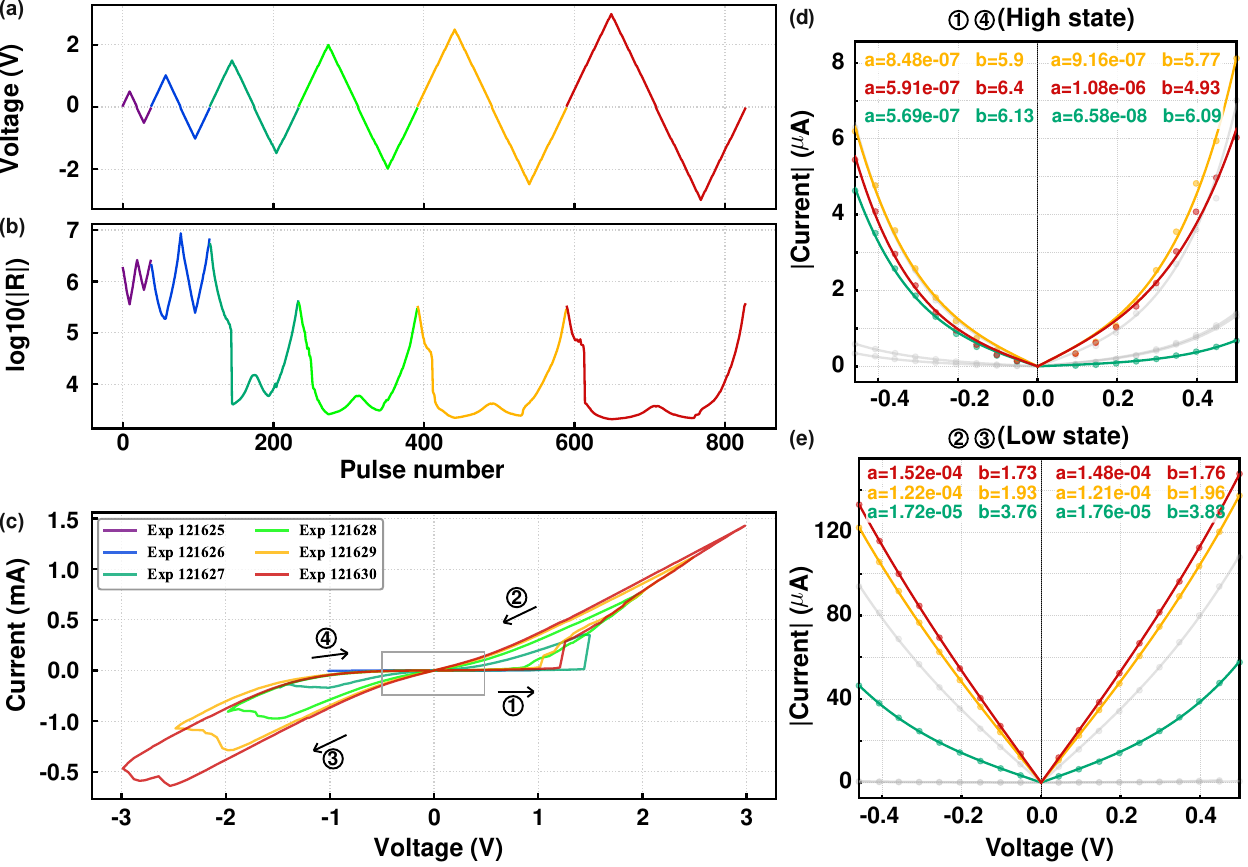}
    \caption{Example hysteresis measurement data and I-V non-linearity feature extraction. Trace colors are consistent across the panels. (a) Applied voltage pulse train, with pulse amplitudes increasing across consecutive \texttt{CT} experiments. (b) Raw resistance evolution plotted against pulse number, with a $\log_{10}$ scale on the y-axis. (c) Hysteresis curves in the I-V plane, with the four sweep branches labeled schematically. (d) Low-voltage sinh fitting for branches 1 and 4, corresponding to the high-resistance readout state. (e) Low-voltage sinh fitting for branches 2 and 3, corresponding to the low-resistance readout state. The fitting parameters $a$ and $b$ are obtained from Equation\,~\ref{eqn:sinh_fit_common}, and the coefficient of determination $R^2$ is used to assess branch-level fit quality. The current magnitude is displayed as $|I|$ for visual comparison across branch polarities.}
    \label{fig:hys_data}
\end{figure}

The I-V non-linearity features describe the branch-level shape of the low-voltage hysteresis response. Figure~\ref{fig:hys_data}(a) shows the applied voltage sequence across consecutive \texttt{CT} experiments, while Figure~\ref{fig:hys_data}(b) shows the corresponding resistance evolution over pulse number. The same measurements are shown in the I-V plane in Figure~\ref{fig:hys_data}(c), where each \texttt{CT} sweep is divided into four branches: $0 \rightarrow V_{\max}$, $V_{\max} \rightarrow 0$, $0 \rightarrow V_{\min}$, and $V_{\min} \rightarrow 0$. Figure~\ref{fig:hys_data}(d,e) then illustrates the low-voltage fitting procedure for the high- and low-resistance branch groups, respectively.

Equation\,~\ref{eqn:sinh_fit_common} in the main text is used here as the reduced fitting form for the low-voltage response of the four \texttt{CT} branches. A sinh-type current-voltage dependence has been used previously in memristor I-V descriptions. In particular, Yang \textit{et al.} used an approximate sinh term to describe ON-state behavior associated with electron tunneling, while Messaris \textit{et al.} discussed a related state-dependent sinh framework \cite{Yang2008Memristive,Messaris2018}. In the present workflow, the same sinh voltage dependence is retained only in a reduced sinh-fitting form for branch-level feature extraction.

Using the Maclaurin series of the hyperbolic sine function~\cite{DLMF_HyperbolicMaclaurin}, Equation\,\ref{eqn:sinh_fit_common} can be expanded as Equation\,\ref{eqn:sinh_maclaurin}:
\begin{equation}
    I(V)
    = a\sinh(bV)
    =
    \underbrace{abV}_{\text{leading linear term}}
    +
    \underbrace{\frac{ab^3V^3}{3!}+\frac{ab^5V^5}{5!}+\frac{ab^7V^7}{7!}+\cdots}_{\text{higher-order non-linear terms}}.
    \label{eqn:sinh_maclaurin}
\end{equation}
Here, the term $abV$ is linear in $V$, whereas the higher-order odd-power terms represent the non-linear contribution. Relative to the leading linear term, this contribution can be written in Equation\,\ref{eqn:sinh_relative_nonlinearity}
\begin{equation}
    \frac{I(V)-abV}{abV}
    =
    \frac{(bV)^2}{3!}
    +
    \frac{(bV)^4}{5!}
    +
    \frac{(bV)^6}{7!}
    +\cdots.
    \label{eqn:sinh_relative_nonlinearity}
\end{equation}
Since the fitting window is restricted to $|V|\leq 0.5$\,V, the following interpretation is limited to the low-voltage regime. Within this window, and for a fixed non-zero voltage magnitude, the relative non-linear contribution increases with $b$. As all retained fitted $b$ values in the present dataset are positive, a larger $b$ is interpreted here as stronger low-voltage non-linearity, whereas a smaller $b$ indicates a response closer to the leading linear term. Accordingly, $b$ is used in the present workflow as a descriptor of low-voltage non-linearity.
\newpage

\subsection{R$_{\text{off}}$/R$_{\text{on}}$ features}
\label{supp-sec:Roff/Ron features}
The R$_{\text{off}}$/R$_{\text{on}}$ features provide a compact operational description of the resistive state changes observed in each hysteresis measurement. Fixed-bias resistance readout is commonly used to compare high- and low-resistance states in resistive-switching devices~\cite{Lanza2019RecommendedMethods,Lanza2021a}.

\begin{figure}[!htbp]
    \centering
    \includegraphics[width=\linewidth]{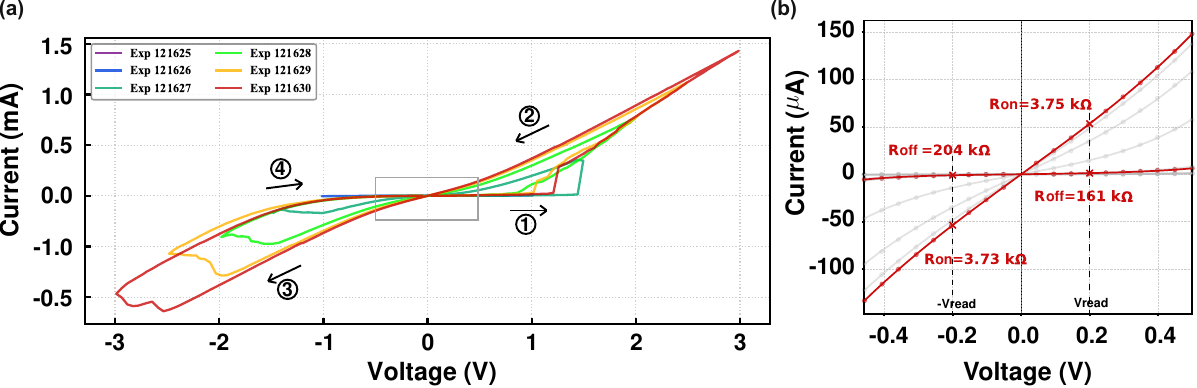}
    \caption{Example hysteresis and R$_{\text{off}}$/R$_{\text{on}}$ features. Trace colors are consistent across the plots. (a) Hysteresis curves in the I-V plane. (b) Example extraction of $R_{\mathrm{on}}$ and $R_{\mathrm{off}}$ using branch-level resistance values at $|V_{\mathrm{read}}|=0.2$\,V.}
    \label{fig:ronroff_data}
\end{figure}
\label{subsec:hys_features_Switching_Window}

Figure~\ref{fig:ronroff_data}(a) shows the four branches used for branch-level resistance readout: branch~1 for \(0 \rightarrow V_{\max}\), branch~2 for \(V_{\max} \rightarrow 0\), branch~3 for \(0 \rightarrow V_{\min}\), and branch~4 for \(V_{\min} \rightarrow 0\). Figure~\ref{fig:r2_combined_report}(b) summarizes the branch-level fit quality using raw resistance values near the selected \(\pm 0.2\)\,V read point. As detailed there, raw points within \([0.195,0.205]\)\,V for positive sweeps and \([-0.205,-0.195]\)\,V for negative sweeps were used for this diagnostic plot. Since not every branch contains a raw point within these read-voltage windows, the final R$_{\text{off}}$/R$_{\text{on}}$ extraction used the fitted sinh curve evaluated at exactly \(\pm 0.2\)\,V for branches with \(R^2 \geq 0.95\).

Each CT sweep was then treated as two polarity-specific branch pairs: branches~1 and~2 for the positive-voltage side, and branches~3 and~4 for the negative-voltage side. Within each pair, the lower resistance at \(|V_{\mathrm{read}}|=0.2\)\,V was assigned as \(R_{\mathrm{on}}\), and the higher resistance was assigned as \(R_{\mathrm{off}}\). The reported ratio was then calculated as R$_{\text{off}}$/R$_{\text{on}}$ for that polarity. Figure~\ref{fig:ronroff_data}(b) shows the zoomed-in low-voltage region, where the current values are reconstructed from the stored voltage and resistance values using \(I=V/R\) for visualization purposes only.

The extracted R$_{\text{off}}$/R$_{\text{on}}$ values should therefore be interpreted as operational electrical descriptors under a fixed read-voltage condition. Large R$_{\text{off}}$/R$_{\text{on}}$ values indicate stronger separation between conductive and resistive readout states, whereas values close to unity indicate weak or negligible separation at $|V_{\mathrm{read}}|=0.2$\,V. Since the calculation depends on specific readout points, the features can be affected by resistance noise, sampling resolution at the read voltage, low-current measurement limits, and branch-to-branch variability. These values are therefore used as descriptive summaries of the measured hysteresis response.
\newpage

\subsection{Switching features}
\label{supp-sec:switching_features}
\begin{figure}[!htbp]
    \centering
    \includegraphics[width=\linewidth]{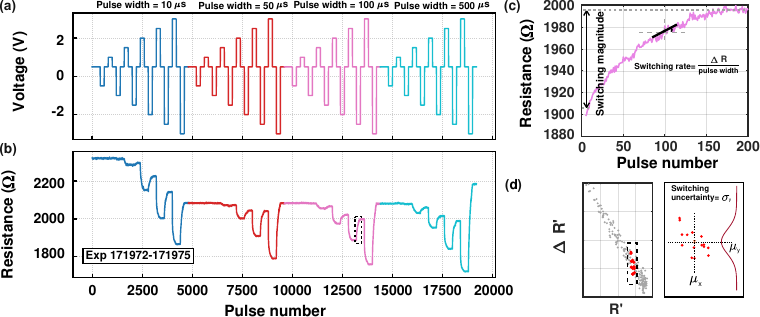}
    \caption{Example resistance switching pulse train, measurement data, and switching-feature extraction. Trace colors are consistent across the panels. (a) Applied voltage pulse train, with pulse widths of $10\,\mu$s, $50\,\mu$s, $100\,\mu$s, and $500\,\mu$s. For each pulse width, the programming voltage is swept from $0.5$\,V to $3.0$\,V in $0.5$\,V steps. (b) Corresponding resistance evolution plotted against pulse number. (c) Example local extraction of switching magnitude and switching rate from the resistance change within a selected pulse window. (d) Schematic of the extraction of switching uncertainty from the local distribution of resistance-change responses, following the switching-uncertainty framework used for memristive devices \cite{Stathopoulos2019SwitchingUncertainty}.}
    \label{fig:pf_data}
\end{figure}

Resistance switching features describe the gradual or abrupt resistance modulation induced by repeated voltage pulses after electroforming. Unlike the electroforming descriptors, which target the first major transition into a conductive state, the switching descriptors summarize how the device resistance evolves under subsequent programming stimuli. Figure~\ref{fig:pf_data}(a) shows the applied pulse sequence, while Figure~\ref{fig:pf_data}(b) shows the corresponding resistance evolution over pulse number. The resistance trajectory is then used to extract local switching descriptors, as illustrated in Figure~\ref{fig:pf_data}(c,d).

In the switching protocol, voltage pulses are applied with pulse widths of $10\,\mu$s, $50\,\mu$s, $100\,\mu$s, and $500\,\mu$s. For each pulse-width block, the programming voltage is swept from $0.5$\,V to $3.0$\,V in $0.5$\,V steps, with repeated pulses applied at each voltage level. The measured resistance sequence, therefore, records the cumulative device response to pulse amplitude, pulse width, pulse polarity, and pulse number. These protocol variables provide the basis for comparing switching behavior across devices and experimental conditions.

The switching magnitude is extracted as the resistance change over a selected local pulse window, denoted here as $\Delta R$. The switching rate is then calculated by normalizing this resistance change by the corresponding pulse width, providing an operational measure of resistance change per unit programming time. Switching uncertainty is estimated from the spread of local switching responses in the resistance-change space, following the idea that memristive switching dynamics are not fully deterministic even under nominally identical stimulation conditions \cite{Stathopoulos2019SwitchingUncertainty}.

The extracted switching features can also be used to construct a switching surface, where the expected resistance change is represented as a function of the current resistance state and pulse stimulus. In this representation, local switching responses are mapped into a state-response space, allowing the average switching trend and its variability to be visualized together. This provides a compact model-oriented summary of how resistance updates depend on the device state and applied pulse conditions, and can support downstream fitting, comparison, and simulation workflows.

The distribution of extracted switching magnitudes should be interpreted cautiously. Large switching magnitudes indicate strong pulse-induced resistance updates, whereas values close to zero indicate weak or negligible change under the selected pulse condition. However, the distribution can be broadened by device-to-device variability, pulse-history dependence, resistance-readout noise, finite sampling resolution, and stochastic switching effects. Therefore, switching magnitude, switching rate, and switching uncertainty are used here as operational summaries for large-scale comparison.
\newpage

\subsection{Volatility features}
\label{supp-sec:volatility_features}
\begin{figure}[!htbp]
    \centering
    \includegraphics[width=\linewidth]{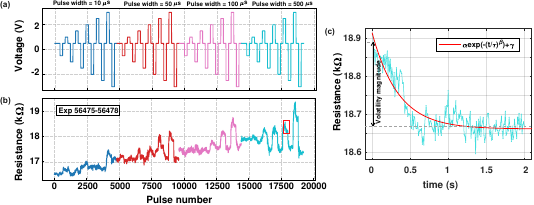}
    \caption{Example resistance switching and volatility measurement data. Trace colors are consistent across the panels. (a) Applied \texttt{PF-IR} voltage sequence, in which each programming segment is followed by an interleaved readout segment. In this example, the post-stimulus readout segment contains 200 read points at $0.5$\,V. (b) Corresponding resistance evolution plotted against pulse number. (c) Example volatility-feature extraction from a post-stimulus readout segment, including the relaxation magnitude and stretched-exponential fitting described by Equation\,~\ref{eqn:volatility_stretched_exp} \cite{Giotis2020BidirectionalVolatile}.}
    \label{fig:volatility_data}
\end{figure}

Volatility features describe the post-stimulus relaxation of device resistance after a programming segment in \texttt{PF-IR} experiments. Unlike the switching features, which summarize resistance changes during active programming, the volatility descriptors are extracted from the interleaved readout segment following each programming-voltage segment. Figure~\ref{fig:volatility_data}(a) shows the combined programming and readout sequence, while Figure~\ref{fig:volatility_data}(b) shows the corresponding resistance evolution over pulse number. Figure~\ref{fig:volatility_data}(c) illustrates the local extraction of volatility features from one post-stimulus readout segment.

For each volatility cycle, the readout segment is fitted using the stretched-exponential form in Equation\,\ref{eqn:volatility_stretched_exp}.
\begin{equation}
    R(t) = \alpha \exp\left[-\left(\frac{t}{\tau}\right)^{\beta}\right] + \gamma,
    \label{eqn:volatility_stretched_exp}
\end{equation}
following the volatility model used for bidirectional volatile signatures in metal-oxide memristors \cite{Giotis2020BidirectionalVolatile}. Here, $t$ is the elapsed time within the readout segment, $\gamma$ is the fitted long-time resistance level, $\alpha$ is the fitted offset from this level at $t=0$, $\tau$ is the relaxation time constant, and $\beta$ is the stretching exponent. The same notation is retained in the deposited database, where the fitted parameters are stored as \texttt{alpha\_ohm}, \texttt{tau\_s}, \texttt{beta}, and \texttt{gamma\_ohm}. The associated readout window and measured resistance states are recorded using \texttt{dt\_s}, \texttt{T\_window\_s}, \texttt{Rpre\_ohm}, \texttt{Rstart\_ohm}, and \texttt{Rend\_ohm}.

The volatility magnitude is interpreted as an operational measure of resistance relaxation over the post-stimulus readout window. In the fitted model, $\alpha$ captures the offset between the initial readout state and the fitted long-time level, while the directly observed finite-window relaxation can be assessed from the difference between \texttt{Rstart\_ohm} and \texttt{Rend\_ohm}. Larger relaxation magnitudes indicate stronger volatile behavior after programming, whereas smaller values indicate more persistent resistance states within the measured time window. These descriptors can be affected by readout noise, finite observation time, pulse-history dependence, and incomplete relaxation within the readout segment. They are therefore used as descriptive summaries of post-stimulus relaxation.
\newpage

\section{Feature statistics summaries, filtering details, and physical explanation}
The following sections detail the statistics of the extracted features, the details of filtering/cleaning procedures, and confirmation of the feature trends against existing understanding in the academic literature.

\subsection{Electroforming distributions and grouping details}
\label{supp-sec:Electroforming distributions and grouping details}

\begin{figure}[!htbp]
    \centering
    \includegraphics[width=0.9\linewidth]{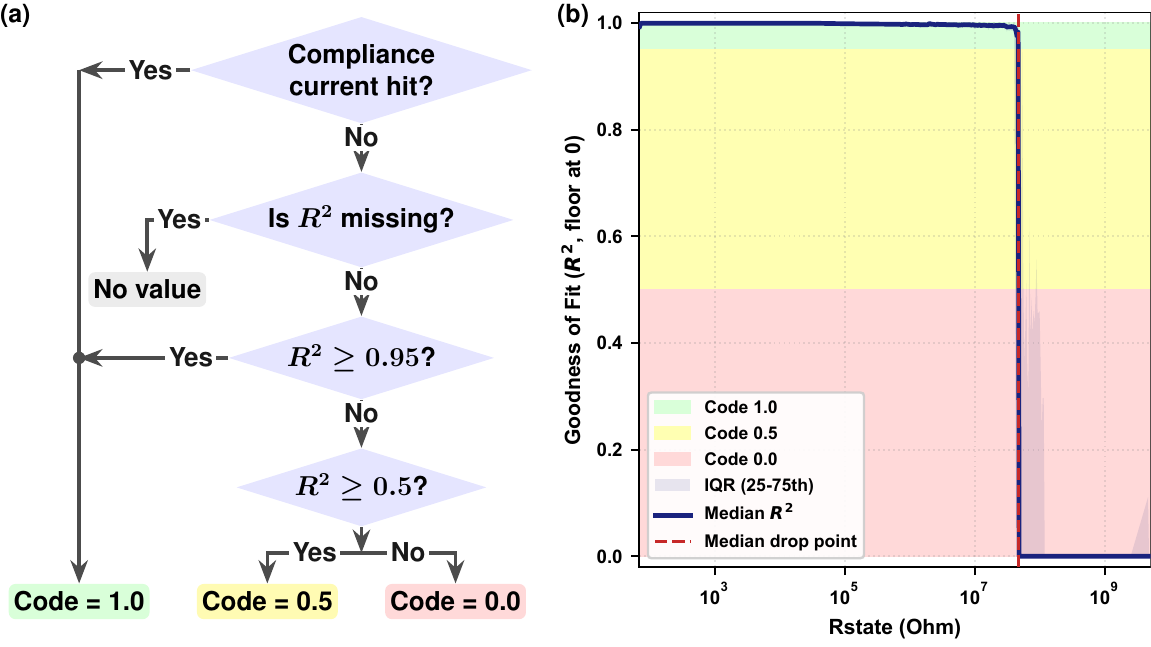}
    \caption{Branch-level coding logic for electroforming classification. 
    (a) Decision tree for assigning branch codes using compliance-current-hit status and sinh-fit $R^2$. Compliance-current hits are assigned Code~1.0; branches without such hits are coded using the $R^2$ thresholds shown. 
    (b) Sinh-fit coefficient of determination ($R^2$) versus $R_{\mathrm{state}}$ at the selected $\pm 0.2$\,V read point for branches without compliance-current hits. Shaded bands indicate the Code~1.0, Code~0.5, and Code~0.0 regions; the solid line and gray band show the median and interquartile range of $R^2$, respectively.}
    \label{fig:r2_combined_report}
\end{figure}

Figure~\ref{fig:r2_combined_report}(a) summarizes the branch-level coding logic used for electroforming classification. The first decision is whether a compliance-current hit is recognized within the low-voltage fitting window ($|V|\leq 0.5$\,V). A branch is marked as having a compliance-current hit when at least two raw points within this window reach the compliance threshold. Such branches are assigned Code~1.0 directly, since a compliance hit already within the low-voltage region indicates a current-limited conductive response. In this case, the branch shape can be truncated by compliance control, so the fitted $R^2$ is not used as the primary decision quantity.

For branches without a recognized low-voltage compliance-current hit, the code is assigned from the coefficient of determination ($R^2$) of the low-voltage sinh fit, $I(V)=a\sinh(bV)$ \cite{Messaris2018,Weisberg2014,Chicco2021}. Branches with missing $R^2$ are treated as having no valid value. The remaining branches are coded as follows: $R^2 \geq 0.95$ gives Code~1.0, $0.5 \leq R^2 < 0.95$ gives Code~0.5, and $R^2 < 0.5$ gives Code~0.0.

Figure~\ref{fig:r2_combined_report}(b) shows the corresponding $R^2$-$R_{\mathrm{state}}$ distribution for branches without compliance-current hits. Here, $R_{\mathrm{state}}$ is taken from the raw point closest to the target read bias, using $[0.195,0.205]$\,V for positive sweeps and $[-0.205,-0.195]$\,V for negative sweeps. Only branches with a raw point inside these read-voltage windows are included in the plot. This retains 428{,}302 branch points from 113{,}980 experiments, corresponding to 98.3\% of the full no-compliance-current-hit branch dataset. Negative $R^2$ values are clipped to zero for visualization.

The trend in Figure~\ref{fig:r2_combined_report}(b) supports using $R^2$ as a practical branch-level indicator for the no-compliance-current-hit cases. The median drop occurs at approximately $R_{\mathrm{state}}\approx 4.73\times10^7~\Omega$: 99.4\% of branches below this point have $R^2\geq 0.95$, whereas 96.3\% of branches above it have $R^2<0.5$. This is consistent with the expectation that high-resistance, low-current pre-forming states are more sensitive to scatter, fluctuations, and low-current acquisition effects \cite{Ielmini2010RTN,Ambrogio2015NoiseBroadening,Lee2023LFNRRAM}. After electroformation, the low-voltage branch response is typically more regular and is more readily captured by the same reduced sinh form.

\begin{table}[!h]
\centering
\caption{Mapping from the final codes assigned to the four \texttt{CurveTracer} branches by the decision tree in Figure~\ref{fig:r2_combined_report}(a) to the experiment-level EF classes, together with experiment counts for each recipe group.}
\label{tab:ef_classification}

\small
\setlength{\tabcolsep}{6.5pt}
\renewcommand{\arraystretch}{1}

\begin{tabular}{l|c|c|c|c}
\hline
\textbf{Classification rule} & \textbf{EF class} & \textbf{HfO$_\text{x}$N$_\text{y}$} & \textbf{TiO$_\text{x}$/AlO$_\text{y}$} & \textbf{TiO$_\text{x}$} \\
\hline

Any branch with final code = 0.5 & \texttt{UNCERTAIN} & 1,723 & 3,995 & 2,242 \\
Code pattern = 0000 & \texttt{NoEF} & 12,727 & 34,844 & 14,598 \\
Code pattern = 0001 & \texttt{NeEF} & 460 & 104 & 737 \\
Code pattern = 0111 & \texttt{PoEF} & 2,097 & 185 & 307 \\
Code pattern = 1111 & \texttt{EF} & 31,867 & 13,784 & 12,106 \\
Any other code pattern & \texttt{OTHER} & 68 & 187 & 100 \\

\hline
\end{tabular}

\vspace{4pt}
\footnotesize
\noindent\textit{Note:} Code order is $0 \rightarrow V_{\max}$, $V_{\max} \rightarrow 0$, $0 \rightarrow V_{\min}$, and $V_{\min} \rightarrow 0$ (positive-up, positive-down, negative-down, and negative-up). \texttt{NoEF}: no electroforming, \texttt{NeEF}: negative voltage electroforming, \texttt{PoEF}: positive voltage electroforming, \texttt{EF}: already electroformed. All counts are experiment-level.
\normalsize
\end{table}

Table~\ref{tab:ef_classification} summarizes the resulting experiment-level EF classes for the subset of CurveTracer experiments in which I-V non-linearity fitting succeeded in all four branches. Across all three recipe groups, \texttt{NoEF} and \texttt{EF} account for most experiments, whereas \texttt{NeEF}, \texttt{PoEF}, and \texttt{OTHER} are comparatively rare. The \texttt{UNCERTAIN} class captures borderline cases. It is assigned whenever any of the four branches receives a final code of 0.5 and is retained for separate examination rather than being forced into a strict EF class.

\begin{figure}[!htbp]
    \centering
    \includegraphics[width=0.8\linewidth]{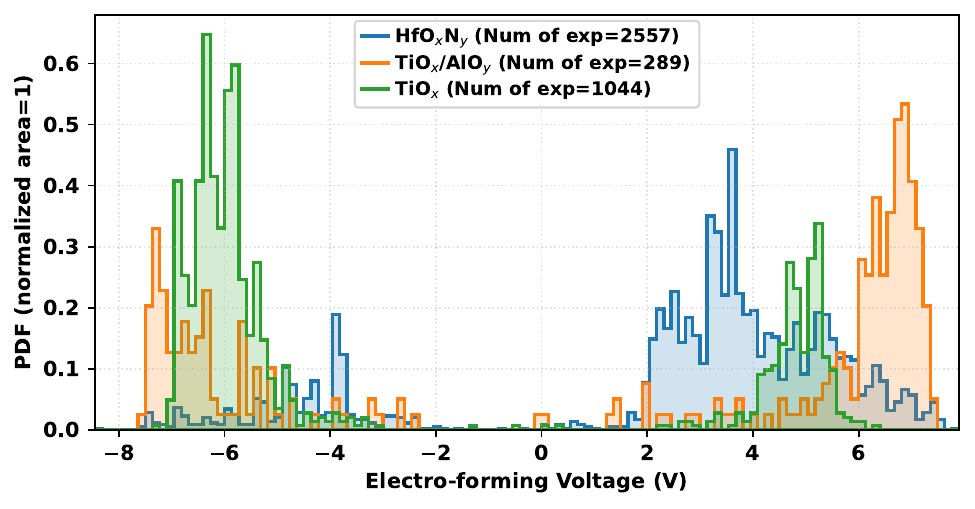}
    \caption{Distribution of electroforming voltage for the HfO$_\text{x}$N$_\text{y}$, TiO$_\text{x}$/AlO$_\text{y}$, and TiO$_\text{x}$ recipe groups. Overlaid histograms are shown for each recipe, and the legend reports the number of experiments in each group.}
    \label{fig:electroform_recipe_distribution}
\end{figure}

Figure~\ref{fig:electroform_recipe_distribution} compares the electroforming-voltage distributions of the three recipe groups after collapsing all available device areas into a single recipe-level distribution. Each histogram is independently normalized to probability density, so the comparison emphasizes the shape and preferred voltage range of each recipe distribution rather than the absolute number of experiments. Positive and negative voltages correspond to positive and negative electroforming events, respectively.

The three recipe groups show clearly different electroforming-voltage distributions. HfO$_\text{x}$N$_\text{y}$ contains 2,557 experiments, of which 2,107 (82.4\%) are positive electroforming events and 450 (17.6\%) are negative electroforming events. Its positive electroforming voltage has a median of 3.72 V with an interquartile range (IQR) of 3.05-5.10 V, while the negative distribution has a median of $-4.32$ V with an IQR of $-5.23$ to $-3.75$ V. TiO$_\text{x}$/AlO$_\text{y}$ contains 289 experiments, with 186 positive events (64.4\%) and 103 negative events (35.6\%). Its positive electroforming voltage is shifted to the highest range among the three recipes, with a median of 6.45 V and an IQR of 5.91-6.82 V, while the negative distribution is centered at a larger magnitude than HfO$_\text{x}$N$_\text{y}$, with a median of $-6.57$ V and an IQR of $-7.11$ to $-5.58$ V. TiO$_\text{x}$ contains 1,044 experiments, but in this case negative electroforming dominates, with 735 negative events (70.4\%) and 309 positive events (29.6\%). Its negative electroforming voltage is concentrated near $-6$ V, with a median of $-5.97$ V and an IQR of $-6.39$ to $-5.69$ V, whereas the positive distribution is centered at a lower voltage than TiO$_\text{x}$/AlO$_\text{y}$, with a median of 4.82 V and an IQR of 4.45-5.11 V. These results show that the electroforming-voltage distribution is strongly recipe-dependent within the dataset analyzed herein.

\begin{figure}[!htbp]
    \centering
    \includegraphics[width=0.7\linewidth]{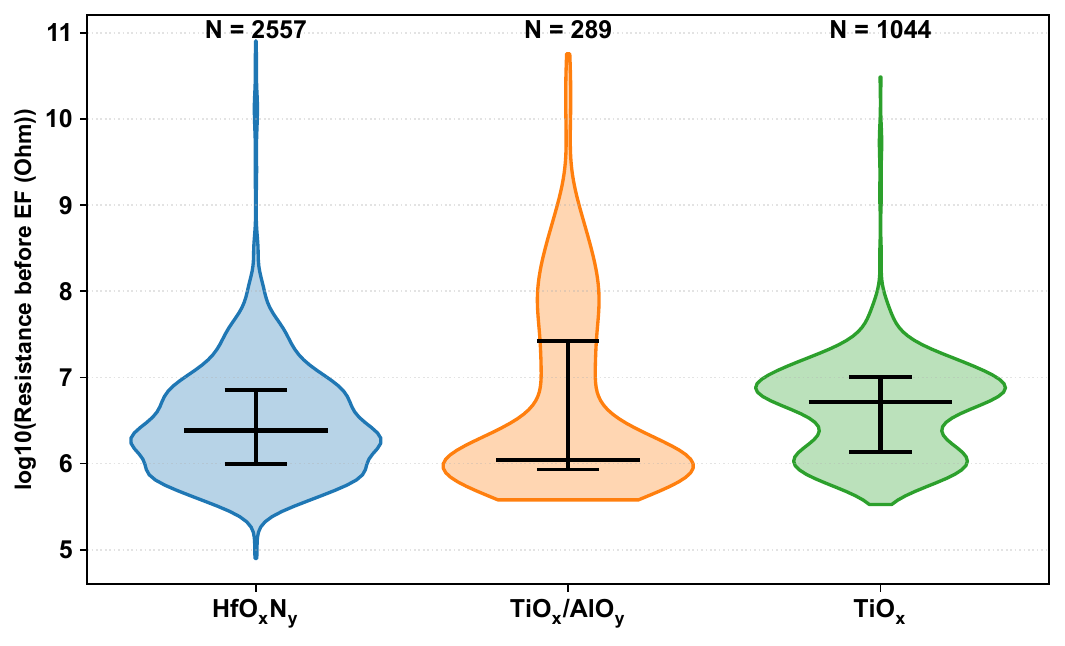}
    \caption{Violin plots of the resistance before electroforming for the HfO$_\text{x}$N$_\text{y}$, TiO$_\text{x}$/AlO$_\text{y}$, and TiO$_\text{x}$ recipe groups. For direct comparison with Figure~\ref{fig:electroform_recipe_distribution}, exactly the same set of experiments is used here, with all available device areas collapsed into a single recipe-level distribution. The vertical axis is shown as $\log_{10}(\mathrm{Resistance\ before\ EF\ (Ohm)})$. The black horizontal bars indicate the 25th percentile, median, and 75th percentile, and the vertical black line spans the interquartile range (IQR). The number of experiments in each recipe group is shown above the corresponding violin.}
    \label{fig:r_before_ohm_violin_by_recipe}
\end{figure}

Using the same experiment set as in Figure~\ref{fig:electroform_recipe_distribution}, Figure~\ref{fig:r_before_ohm_violin_by_recipe} shows the distribution of resistance before electroforming for the three recipe groups, with all available device areas collapsed into a single recipe-level distribution. The pre-electroforming resistance for each experiment was taken from the extracted resistance value immediately before electroforming. Since the resistance spans several orders of magnitude, it is plotted on a log$_{10}$ scale. The three recipe groups show clear differences in both their median resistance and their distribution width.

HfO$_\text{x}$N$_\text{y}$ contains 2,557 experiments. Its median resistance before electroforming is 2.42 M$\Omega$, with an interquartile range (IQR) of 0.99-7.16 M$\Omega$. TiO$_\text{x}$/AlO$_\text{y}$ contains 289 experiments and shows the lowest median resistance but the broadest upper spread, with a median of 1.10 M$\Omega$ and an IQR of 0.86-26.45 M$\Omega$. TiO$_\text{x}$ contains 1,044 experiments and shows the highest median resistance before electroforming, with a median of 5.22 M$\Omega$ and an IQR of 1.37-10.11 M$\Omega$.

These results indicate that the resistance before electroforming is strongly recipe-dependent in the present dataset. In particular, TiO$_\text{x}$/AlO$_\text{y}$ shows the largest variability, while TiO$_\text{x}$ is shifted toward higher pre-electroforming resistance. HfO$_\text{x}$N$_\text{y}$ lies between these two cases, suggesting that the initial resistance state prior to electroforming is systematically influenced by the fabrication recipe.

\begin{figure}[!htbp]
    \centering
    \includegraphics[width=0.7\linewidth]{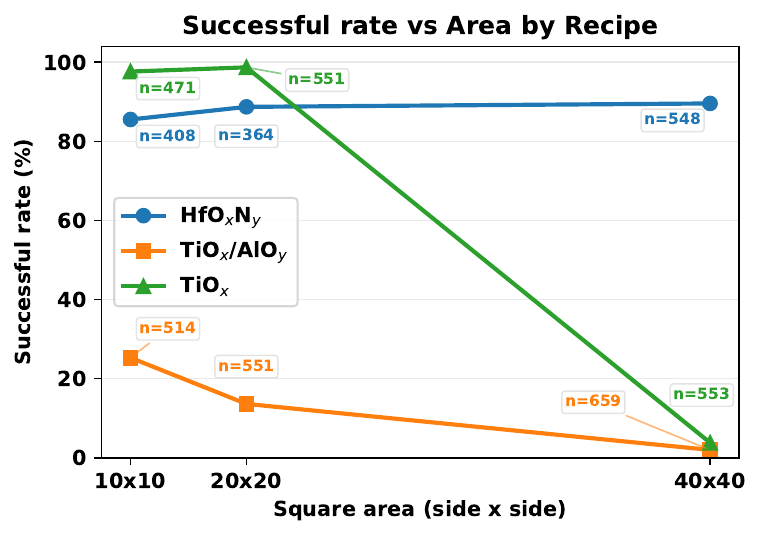}
    \caption{Electroforming success rates for different devices across three device areas for the three recipe groups. Labels indicate the number of devices ($n$) used to determine the success rate for each group.}
    \label{fig:ef_success_rate_vs_area_by_recipe}
\end{figure}

Table~\ref{tab:ef_classification} reports the electroforming category classification of each individual experiment and Figure~\ref{fig:ef_success_rate_vs_area_by_recipe} summarizes the outcome at the device level by following the ordered EF-class sequence of each device across repeated experiments. In this step, experiment-level labels \texttt{UNCERTAIN} and \texttt{OTHER} are not used to define the device-level forming outcome; only \texttt{NoEF}, \texttt{NeEF}, \texttt{PoEF}, and \texttt{EF} are retained in the sequence.

A device was counted as \texttt{NEVER\_FORMED} if all retained experiments were \texttt{NoEF}. A device was counted as \texttt{SUCCESS\_FORMED} if the retained sequence started from \texttt{NoEF}, passed through at least one intermediate state (\texttt{NeEF} or \texttt{PoEF}), and then reached \texttt{EF} and remained \texttt{EF} thereafter. A device was counted as \texttt{PRE\_FORMED} if its retained experiments were already \texttt{EF} from the beginning, meaning that the device was already in a formed state before the observed sequence. Sequences that did not match either the \texttt{NEVER\_FORMED} or \texttt{SUCCESS\_FORMED} pattern were treated as irregular and were not included in the success-rate calculation.

The device-level electroforming success rate was then calculated as the number of \texttt{SUCCESS\_FORMED} devices divided by the sum of \texttt{SUCCESS\_FORMED} and \texttt{NEVER\_FORMED} devices.

Figure~\ref{fig:ef_success_rate_vs_area_by_recipe} shows that the device-level success rate depends strongly on both recipe and device area. HfO$_\text{x}$N$_\text{y}$ maintains a relatively high success rate across all three areas, increasing from 85.5\% at $10\,\mu m\times10\,\mu m$ to 89.6\% at $40\,\mu m\times40\,\mu m$. TiO$_\text{x}$/AlO$_\text{y}$ shows a much lower success rate and decreases monotonically with area, from 25.3\% to 2.0\%. TiO$_\text{x}$ exhibits the highest success rate at smaller areas, with 97.7\% at $10\times10$ and 98.7\% at $20\,\mu m\times20\,\mu m$, but drops sharply to 3.8\% at $40\,\mu m\times40\,\mu m$. These results indicate that electroforming success is not governed by a single experiment-level rule alone, but also depends on the device stack and area when viewed at the device level.
\newpage

\subsection{Low-bias non-linearity summaries}
\label{supp-sec:Low-bias non-linearity summaries}

\begin{figure}[!htbp]
    \centering
    \includegraphics[width=1\linewidth]{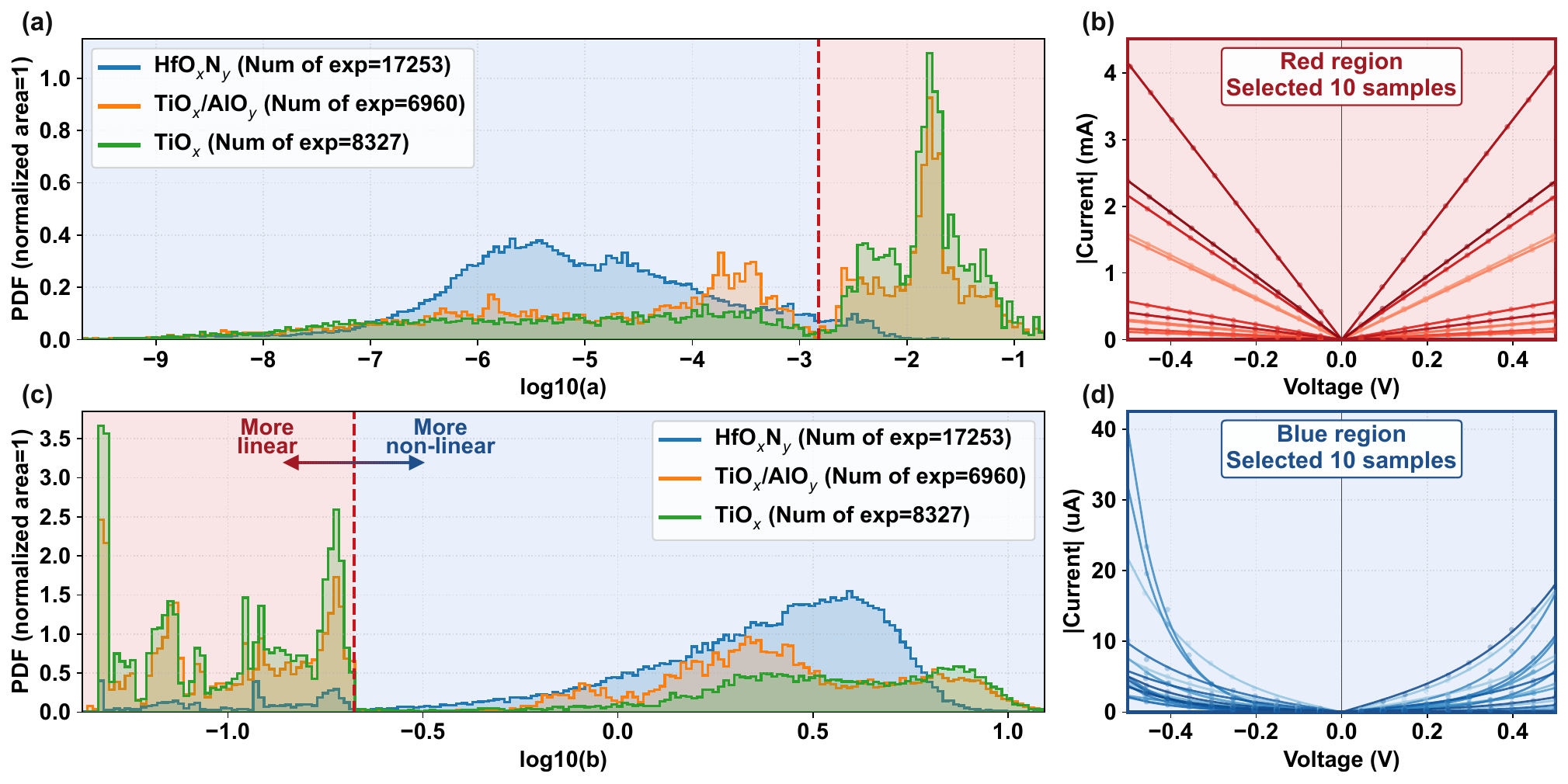}
    \caption{Recipe-wise distributions of the fitted $sinh$ parameters $a$ and $b$ for electroformed experiments without compliance-current hit. Panels (a) and (c) show $\log_{10}(a)$ and $\log_{10}(b)$, respectively, and panels (b) and (d) show illustrative fitted $I$-$V$ branches from the two highlighted overlap regions.}
    \label{fig:iv_nonlinearity_fitting_distribution}
\end{figure}

Figure~\ref{fig:iv_nonlinearity_fitting_distribution} summarizes the fitted $sinh$ parameters for electroformed experiments without compliance-current hit in the three recipe groups, with each recipe histogram independently normalized to unit area. The dataset contains 17,253 HfO$_\text{x}$N$_\text{y}$ experiments, 6,960 TiO$_\text{x}$/AlO$_\text{y}$ experiments, and 8,327 TiO$_\text{x}$ experiments. Since $b$ is used in the present workflow as a practical descriptor of low-voltage I-V non-linearity, the discussion below focuses on panel~(c).

Panel~(c) shows that $\log_{10}(b)$ differs systematically across recipes, with medians of 0.405 for HfO$_\text{x}$N$_\text{y}$, $-0.068$ for TiO$_\text{x}$/AlO$_\text{y}$, and $-0.742$ for TiO$_\text{x}$. Consistent with the interpretation in Section~\ref{supp-sec:Hysteresis features: I-V non-linearity}, larger $b$ corresponds to stronger low-voltage non-linearity within the present fitting window, whereas smaller $b$ corresponds to a response closer to the leading linear term. The empirical split at $\log_{10}(b)=-0.677$ is therefore used here to distinguish a lower-$b$ population, whose fitted branches remain more nearly linear near low bias, from a higher-$b$ population with stronger low-bias curvature. Relative to this split, only 6.7\% of HfO$_\text{x}$N$_\text{y}$ branch values fall in the lower-$b$ region, compared with 45.5\% for TiO$_\text{x}$/AlO$_\text{y}$ and 60.7\% for TiO$_\text{x}$.

Panels~(b) and (d) provide illustrative fitted $I$-$V$ branches from two overlap regions defined jointly by the $a$ and $b$ splits, rather than the basis of the statistical distribution itself. Even so, they remain visually consistent with the separation in panel~(c): the blue-region examples show stronger low-bias curvature than the red-region examples. These results indicate clear recipe-dependent differences in low-voltage I-V non-linearity.
\newpage

\subsection{R$_{\text{off}}$/R$_{\text{on}}$ subset analysis}
\label{supp-sec:Roff/Ron subset analysis}

\begin{figure}[!htbp]
    \centering
    \includegraphics[width=0.75\linewidth]{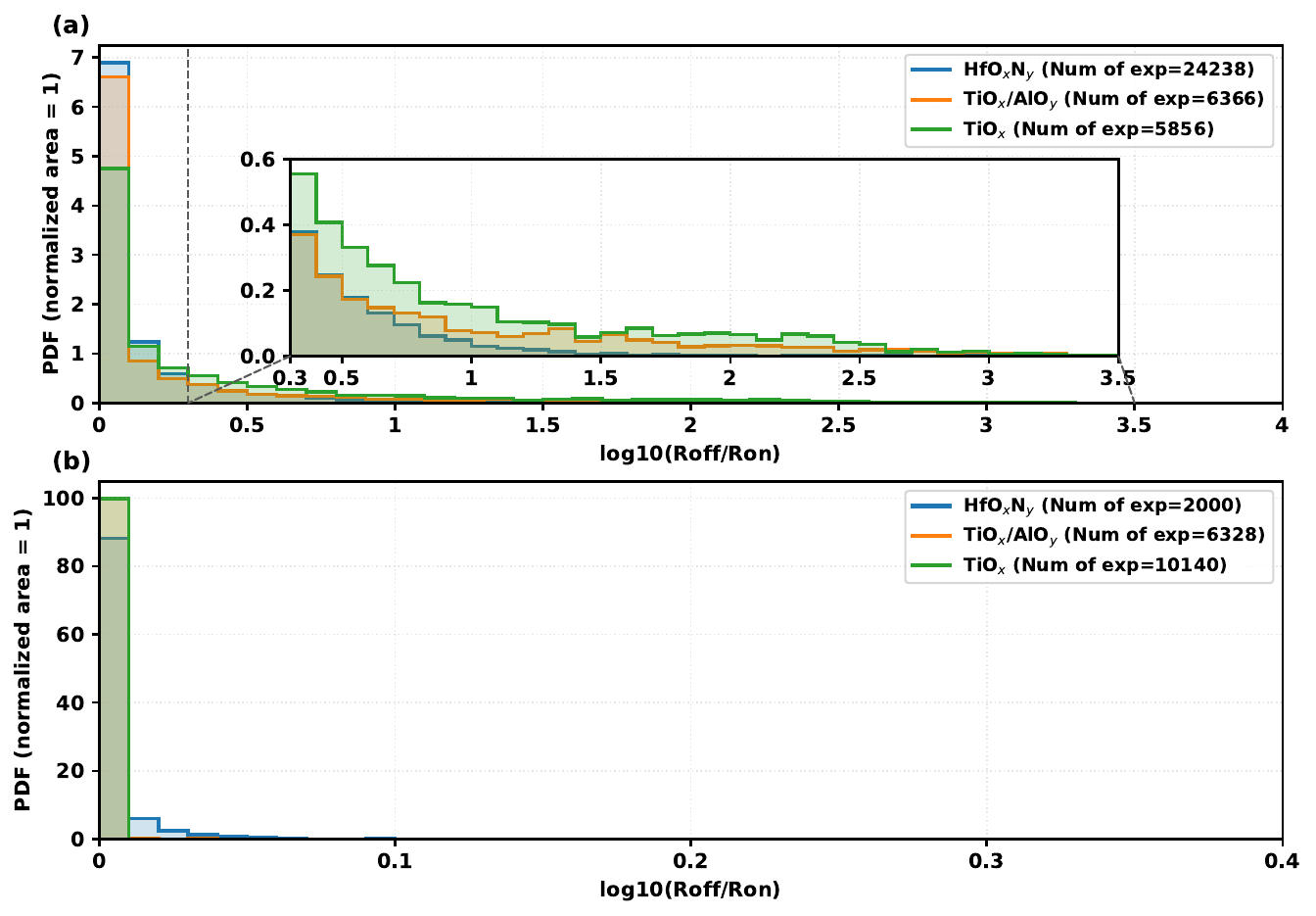}
    \caption{Probability-density distributions of $\log_{10}(R_\mathrm{off}/R_\mathrm{on})$ across recipes after dividing experiments by the empirical mean-$b$ split at $\log_{10}(b)=-0.677$ introduced from Figure~\ref{fig:iv_nonlinearity_fitting_distribution}. Panel (a) shows the higher-$b$ subset and panel (b) the lower-$b$ subset.}
    \label{fig:roff_ron_bsplit_combined_distribution}
\end{figure}

Figure~\ref{fig:roff_ron_bsplit_combined_distribution} uses the mean-$b$ split introduced from Figure~\ref{fig:iv_nonlinearity_fitting_distribution}. For each experiment, the four fitted $\log_{10}(b)$ values are averaged; experiments with mean $\log_{10}(b)>-0.677$ are assigned to the higher-$b$ subset, and those with mean $\log_{10}(b)\leq -0.677$ to the lower-$b$ subset. Only experiments with valid $R_\mathrm{off}/R_\mathrm{on}$ extraction and zero compliance current on both polarities are retained, and branches with $\left|V_\mathrm{max}\right|>3$~V are excluded before pooling the positive- and negative-polarity branches.

Panel (a) uses 40 bins over $0\leq \log_{10}(R_\mathrm{off}/R_\mathrm{on})\leq 4$, whereas panel (b) uses 40 bins over $0\leq \log_{10}(R_\mathrm{off}/R_\mathrm{on})\leq 0.4$. The inset in panel (a) is a zoom of the same pooled histogram, restricted to $0.3\leq \log_{10}(R_\mathrm{off}/R_\mathrm{on})\leq 3.5$. The upper limit in panel (a) excludes only 3 of the 36,460 pooled higher-$b$ branch values (0.008\%), and the truncation in panel (b) excludes 7 of the 18,468 pooled lower-$b$ branch values (0.038\%); both panels therefore retain all data in their respective subsets essentially.

The higher-$b$ subset [Figure~\ref{fig:roff_ron_bsplit_combined_distribution}(a)] retains clear recipe-dependent structure. HfO$_\text{x}$N$_\text{y}$ and TiO$_\text{x}$/AlO$_\text{y}$ remain concentrated at relatively small $\log_{10}(R_\mathrm{off}/R_\mathrm{on})$ values, whereas TiO$_\text{x}$ is broader, more strongly right-shifted, and shows the heaviest upper tail. In pooled branch-level terms, the higher-$b$ subset has a median $\log_{10}(R_\mathrm{off}/R_\mathrm{on})$ of 0.0386, an interquartile range from 0.0051 to 0.1816, and a 99th percentile of 2.077.

The inset in Figure~\ref{fig:roff_ron_bsplit_combined_distribution}(a) shows the upper-tail part of the pooled higher-$b$ distribution more clearly, which is compressed in the main panel by the strong peak near $R_\mathrm{off}/R_\mathrm{on}=1$. Within this filtered subset and restricted range, TiO$_\text{x}$ remains the broadest and most right-shifted of the three recipe distributions. In particular, the number of branches satisfying $R_\mathrm{off}/R_\mathrm{on}\geq 2$ is 3055/24238 (12.60\%) for HfO$_\text{x}$N$_\text{y}$, 1295/6366 (20.34\%) for TiO$_\text{x}$/AlO$_\text{y}$, and 1979/5856 (33.79\%) for TiO$_\text{x}$. For the stricter threshold $R_\mathrm{off}/R_\mathrm{on}\geq 10$, the corresponding values are 303/24238 (1.25\%), 493/6366 (7.74\%), and 745/5856 (12.72\%), respectively.

The lower-$b$ subset [Figure~\ref{fig:roff_ron_bsplit_combined_distribution}(b)] remains dominated by near-unity $R_\mathrm{off}/R_\mathrm{on}$ values. Its pooled branch-level median $\log_{10}(R_\mathrm{off}/R_\mathrm{on})$ is $1.69\times10^{-4}$, with an interquartile range from $7.8\times10^{-5}$ to $3.20\times10^{-4}$ and a 99th percentile of 0.0158. Thus, the lower-$b$ subset remains concentrated very close to $R_\mathrm{off}/R_\mathrm{on}=1$ for the overwhelming majority of branches, although rare larger events are still present.

The mean-$b$ split at $\log_{10}(b)=-0.677$ remains useful as an empirical supplementary-comparison boundary. The higher-$b$ subset retains clearer recipe-dependent $R_\mathrm{off}/R_\mathrm{on}$ structure and a broader upper tail, whereas the lower-$b$ subset remains concentrated near unity for almost all branches.

\begin{figure}[!htbp]
    \centering
    \includegraphics[width=1\linewidth]{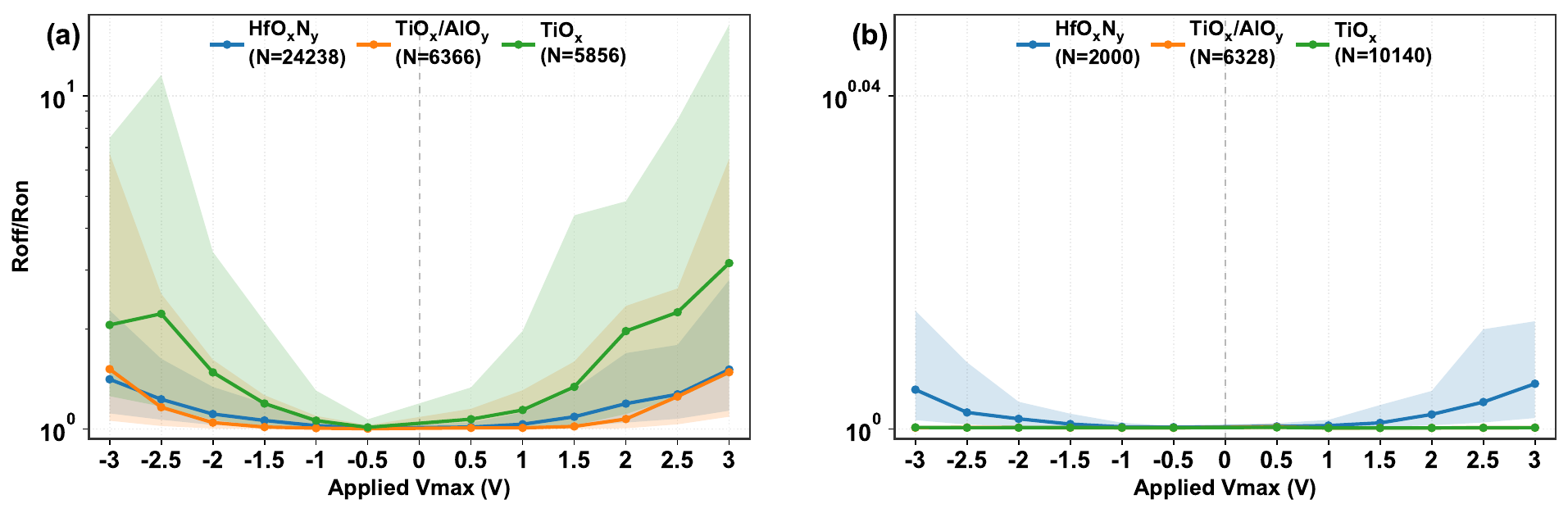}
    \caption{Median $R_\mathrm{off}/R_\mathrm{on}$ as a function of applied $V_\mathrm{max}$ after dividing experiments by the empirical mean-$b$ split at $\log_{10}(b)=-0.677$. Here, the split is applied at the experiment level using the mean of the four fitted branch values of $\log_{10}(b)$. Panel (a) shows the higher-$b$ subset and panel (b) the lower-$b$ subset. Solid lines indicate the median, and shaded bands indicate the interquartile range (IQR).}
    \label{fig:roff_ron_vmax_meanb_ab}
\end{figure}

Figure~\ref{fig:roff_ron_vmax_meanb_ab} uses the same experiment-level mean-$b$ grouping as Figure~\ref{fig:iv_nonlinearity_fitting_distribution} and as the pooled distribution comparison in Figure~\ref{fig:roff_ron_bsplit_combined_distribution}. Experiments were assigned to the higher-$b$ or lower-$b$ subset using the mean of the fitted positive-up, positive-down, negative-down, and negative-up branch values of $\log_{10}(b)$, with the empirical boundary at $-0.677$. On the R$_{\text{off}}$/R$_{\text{on}}$ side, the analysis was restricted to rows with non-null positive and negative $V_\mathrm{max}$, non-null positive and negative R$_{\text{off}}$/R$_{\text{on}}$ values, non-null branch labels on both polarities, and compliance-current limit disabled on both polarities. Branches with $\left|V_\mathrm{max}\right|>3$~V were then excluded branch-wise from the corresponding polarity side.

The two subsets show clearly different voltage dependence. In the higher-$b$ subset [Figure~\ref{fig:roff_ron_vmax_meanb_ab}(a)], $R_\mathrm{off}/R_\mathrm{on}$ remains close to unity at small $\left|V_\mathrm{max}\right|$ but increases at larger voltage magnitude, with clear recipe dependence. TiO$_\text{x}$ shows the strongest increase and the broadest IQR: its median $R_\mathrm{off}/R_\mathrm{on}$ rises from 1.01 at $-0.5$~V and 1.07 at $+0.5$~V to 2.05 at $-3.0$~V and 3.15 at $+3.0$~V. Over the same range, HfO$_\text{x}$N$_\text{y}$ and TiO$_\text{x}$/AlO$_\text{y}$ increase more moderately, with medians of about 1.41 and 1.51 at $-3.0$~V and about 1.51 and 1.48 at $+3.0$~V, respectively. The IQR at larger $\left|V_\mathrm{max}\right|$ further indicates that broader R$_{\text{off}}$/R$_{\text{on}}$ responses are retained in the higher-$b$ population.

In contrast, the lower-$b$ subset [Figure~\ref{fig:roff_ron_vmax_meanb_ab}(b)] remains tightly concentrated near $R_\mathrm{off}/R_\mathrm{on}=1$ across the full voltage range. For all three recipe groups, the medians stay very close to unity even at $\pm 3.0$~V, and the IQRs remain narrow. At $+3.0$~V, the medians are 1.013 for HfO$_\text{x}$N$_\text{y}$, 1.0004 for TiO$_\text{x}$/AlO$_\text{y}$, and 1.0004 for TiO$_\text{x}$; at $-3.0$~V, they are 1.011, 1.0004, and 1.0004, respectively. Thus, even under the less stringent mean-based grouping, the lower-$b$ subset remains dominated by near-unity $R_\mathrm{off}/R_\mathrm{on}$, whereas the higher-$b$ subset retains clearer recipe-dependent structure and stronger voltage dependence.

Overall, the empirical mean-$b$ split at $\log_{10}(b)=-0.677$ still separates a higher-$b$ population with clearer $R_\mathrm{off}/R_\mathrm{on}$ structure and stronger voltage dependence from a lower-$b$ population concentrated near unity, without implying that this compliance-current limit represents a universal physical threshold.
\newpage

\subsection{Switching magnitude summaries}
\label{supp-sec:Switching magnitude summaries}

\begin{figure}[h!]
    \centering
    \includegraphics[width=0.7\linewidth]{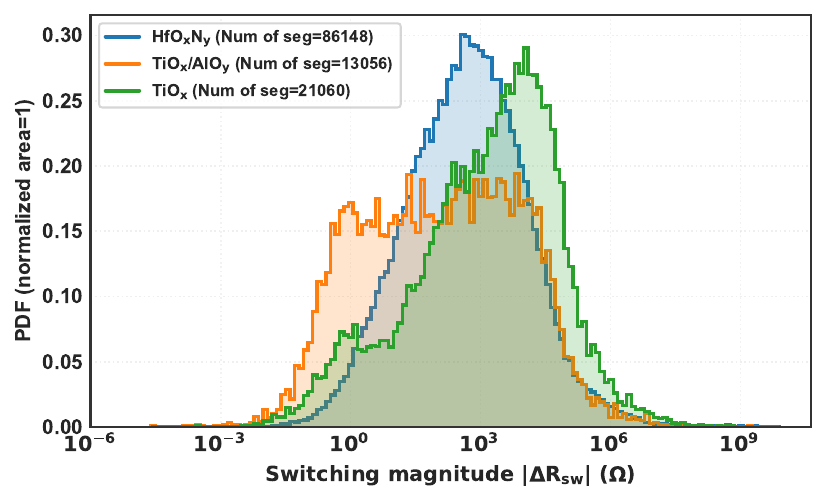}
    \caption{Distribution of absolute switching magnitude, $|\Delta R_{\mathrm{sw}}|$ defined in Equation\,~\ref{eqn:switching_delta}, for the HfO$_\text{x}$N$_\text{y}$, TiO$_\text{x}$/AlO$_\text{y}$, and TiO$_\text{x}$ recipe groups. The horizontal axis is logarithmic, and each probability density function is normalized to unit area. The legend gives the number of retained switching segments in each recipe group.}
    \label{fig:switching_pdf_recipe_pubquality}
\end{figure}

Figure~\ref{fig:switching_pdf_recipe_pubquality} compares the overall distributions of absolute switching magnitude for the three recipe groups in the filtered PF-IR dataset. This analysis included PF-IR experiments with valid structure parameters. The dataset was further restricted to cases for which the nearest previous CT was assigned to the electroformed classes \texttt{NeEF}, \texttt{PoEF}, or \texttt{EF} (Table~\ref{tab:ef_classification}). For each switching segment,
\begin{equation}
    \Delta R_{\mathrm{sw}} = R_{\mathrm{sw,end}} - R_{\mathrm{sw,start}}
    \label{eqn:switching_delta}
\end{equation}
with $R_{\mathrm{sw,start}}$ and $R_{\mathrm{sw,end}}$ estimated from the mean of the first and last 10 points, respectively. In the final exclusion step, entire experiments were removed only when all switching segments within the experiment satisfied $|\Delta R_{\mathrm{sw}}|<10~\Omega$. 

The three recipe groups show clear differences in the overall distribution of $|\Delta R_{\mathrm{sw}}|$. After filtering, the retained dataset contains 86,148 switching segments for HfO$_\text{x}$N$_\text{y}$, 13,056 for TiO$_\text{x}$/AlO$_\text{y}$, and 21,060 for TiO$_\text{x}$, with corresponding median $|\Delta R_{\mathrm{sw}}|$ values of 497.2~$\Omega$, 134.8~$\Omega$, and 2.22~k$\Omega$. This places TiO$_\text{x}$/AlO$_\text{y}$ at the lower-switching end of the comparison, HfO$_\text{x}$N$_\text{y}$ in an intermediate position, and TiO$_\text{x}$ at the higher-switching end. The same ordering is also evident in the upper tail of the distributions: TiO$_\text{x}$ reaches substantially larger values at both the 95th and 99th percentiles ($p_{95}=287.6$~k$\Omega$, $p_{99}=3.95$~M$\Omega$) than HfO$_\text{x}$N$_\text{y}$ ($p_{95}=73.5$~k$\Omega$, $p_{99}=1.24$~M$\Omega$) and TiO$_\text{x}$/AlO$_\text{y}$ ($p_{95}=66.5$~k$\Omega$, $p_{99}=0.913$~M$\Omega$). Even so, the three distributions are not fully separated and still overlap visibly, particularly in the approximate $10^2$-$10^4~\Omega$ range.
\begin{figure}[!htbp]
    \centering
    \includegraphics[width=0.7\linewidth]{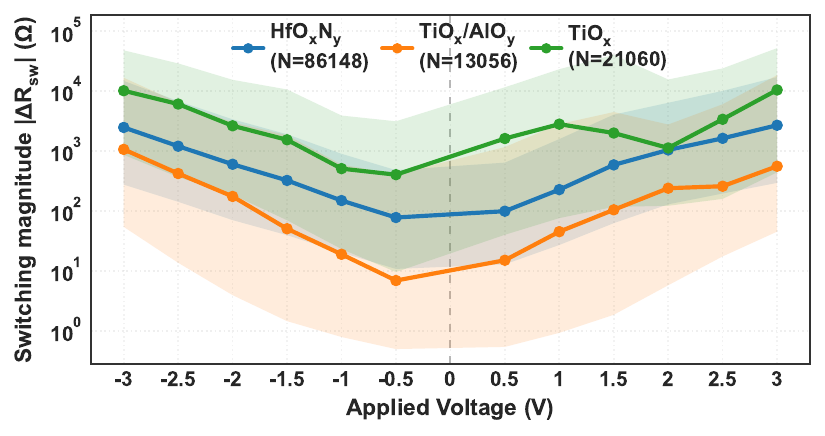}
    \caption{Absolute switching magnitude, $|\Delta R_{\mathrm{sw}}|$, as a function of applied voltage for the HfO$_\text{x}$N$_\text{y}$, TiO$_\text{x}$/AlO$_\text{y}$, and TiO$_\text{x}$ recipe groups, using the same filtered PF-IR dataset as in Figure~\ref{fig:switching_pdf_recipe_pubquality}. Markers show the median, shaded regions show the interquartile range, and the vertical axis is logarithmic.}
    \label{fig:switching_pdf_voltage_segment_pubquality_abs10_trial_narrow_n}
\end{figure}

Figure~\ref{fig:switching_pdf_voltage_segment_pubquality_abs10_trial_narrow_n} shows the same filtered switching dataset as Figure~\ref{fig:switching_pdf_recipe_pubquality}, but separated by applied voltage rather than combined into a single recipe-level distribution.  In all three recipe groups, the median $|\Delta R_{\mathrm{sw}}|$ is smallest at low $|V|$ and tends to increase as the voltage magnitude becomes larger, giving a broad V-shaped dependence. This behavior is not perfectly symmetric between positive and negative bias, and it is not strictly monotonic at every voltage step.

Across the full voltage range, TiO$_\text{x}$/AlO$_\text{y}$ remains the lowest-switching recipe, HfO$_\text{x}$N$_\text{y}$ remains intermediate, and TiO$_\text{x}$ is generally the highest-switching recipe and shows the broadest spread. HfO$_\text{x}$N$_\text{y}$ follows a clear V-shaped trend, increasing from 78.3~$\Omega$ at $-0.5$~V to 2475.7~$\Omega$ at $-3$~V and 2692.6~$\Omega$ at $+3$~V, with slightly larger medians on the positive side at matched $|V|$. TiO$_\text{x}$/AlO$_\text{y}$ also increases away from low voltage, from 7.0~$\Omega$ at $-0.5$~V to 1067.7~$\Omega$ at $-3$~V and 558.9~$\Omega$ at $+3$~V, indicating clear polarity asymmetry. TiO$_\text{x}$ remains at the high-switching end throughout the voltage range, with a median of 404.5~$\Omega$ at $-0.5$~V and about 10.1-10.4~k$\Omega$ at $\pm 3$~V, although the positive-bias branch is not strictly monotonic between $+1$ and $+2$~V. Taken together, these results show that the recipe ordering seen in Figure~\ref{fig:switching_pdf_recipe_pubquality} is maintained after separating the data by voltage, while also revealing a clear dependence on bias magnitude and polarity.
\newpage

\subsection{Volatility magnitude summaries}
\label{supp-sec:Volatility magnitude summaries}

\begin{figure}[h!]
    \centering
    \includegraphics[width=0.75\linewidth]{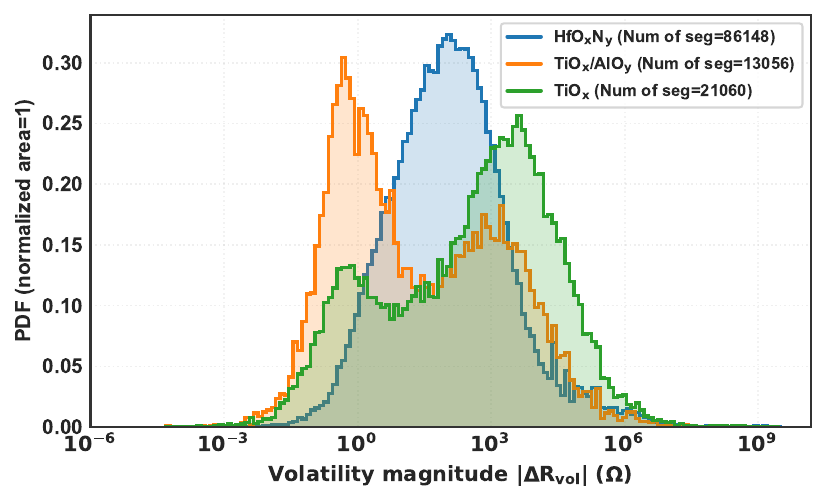}
    \caption{Distribution of absolute volatility magnitude, $|\Delta R_{\mathrm{vol}}|$ defined in Equation\,~\ref{eqn:volatility_delta}, for the HfO$_\text{x}$N$_\text{y}$, TiO$_\text{x}$/AlO$_\text{y}$, and TiO$_\text{x}$ recipe groups. The horizontal axis is logarithmic, and each probability density function is normalized to unit area. The legend gives the number of retained segments in each recipe group.}
    \label{fig:volatility_pdf_recipe_pubquality_excluding_allseg_abs_sw_lt10}
\end{figure}

Figure~\ref{fig:volatility_pdf_recipe_pubquality_excluding_allseg_abs_sw_lt10} compares the overall distributions of absolute volatility magnitude for the three recipe groups in the filtered PF-IR dataset. This analysis included PF-IR experiments with valid structure parameters. The dataset was further restricted to cases for which the device was assigned to one of the electroformed classes \texttt{NeEF}, \texttt{PoEF}, or \texttt{EF} (Table~\ref{tab:ef_classification}). For each volatility segment,
\begin{equation}
    \Delta R_{\mathrm{vol}} = R_{\mathrm{vol,end}} - R_{\mathrm{vol,start}}
    \label{eqn:volatility_delta}
\end{equation}
with $R_{\mathrm{vol,start}}$ and $R_{\mathrm{vol,end}}$ estimated from the mean of the first and last 10 points, respectively. To retain only experiments with measurable switching activity, entire experiments were removed when all switching segments within the experiment satisfied $|\Delta R_{\mathrm{sw}}|<10~\Omega$.

The three recipe groups show clear differences in the overall distribution of $|\Delta R_{\mathrm{vol}}|$. After filtering, the retained dataset contains 86,148 segments for HfO$_\text{x}$N$_\text{y}$, 13,056 for TiO$_\text{x}$/AlO$_\text{y}$, and 21,060 for TiO$_\text{x}$, with corresponding median $|\Delta R_{\mathrm{vol}}|$ values of 107.5~$\Omega$, 10.64~$\Omega$, and 647.7~$\Omega$. This places TiO$_\text{x}$/AlO$_\text{y}$ at the lower-volatility end of the comparison, HfO$_\text{x}$N$_\text{y}$ in an intermediate position, and TiO$_\text{x}$ at the higher-volatility end. TiO$_\text{x}$ also shows the strongest upper tail, with $p_{95}=153.9$~k$\Omega$ and $p_{99}=1.51$~M$\Omega$. HfO$_\text{x}$N$_\text{y}$ remains intermediate overall but also retains a pronounced long tail ($p_{95}=27.5$~k$\Omega$, $p_{99}=970.7$~k$\Omega$), whereas TiO$_\text{x}$/AlO$_\text{y}$ remains lower overall despite partial overlap in the upper tail ($p_{95}=28.2$~k$\Omega$, $p_{99}=380.0$~k$\Omega$).

\begin{figure}[!htbp]
    \centering
    \includegraphics[width=0.75\linewidth]{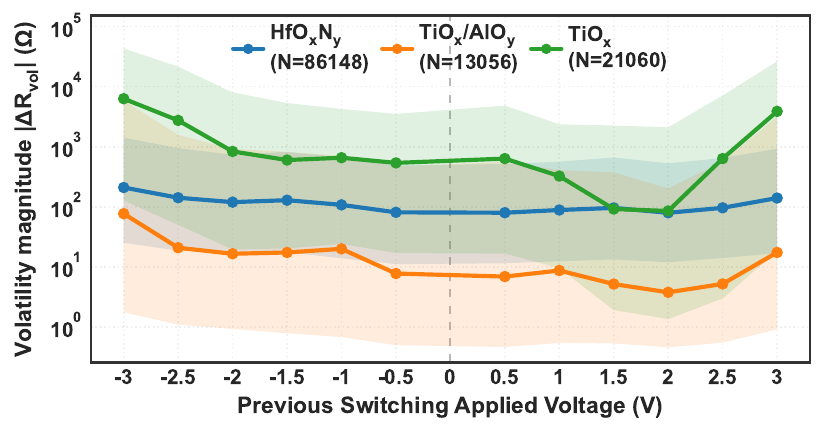}
    \caption{Absolute volatility magnitude, $|\Delta R_{\mathrm{vol}}|$, as a function of the preceding switching voltage for the HfO$_\text{x}$N$_\text{y}$, TiO$_\text{x}$/AlO$_\text{y}$, and TiO$_\text{x}$ recipe groups, using the same filtered PF-IR dataset as in Figure~\ref{fig:volatility_pdf_recipe_pubquality_excluding_allseg_abs_sw_lt10}. Markers show the median, shaded regions show the interquartile range, and the vertical axis is logarithmic.}
    \label{fig:volatility_pdf_voltage_segment_pubquality_abs10}
\end{figure}

Figure~\ref{fig:volatility_pdf_voltage_segment_pubquality_abs10} shows the same filtered volatility dataset as Figure~\ref{fig:volatility_pdf_recipe_pubquality_excluding_allseg_abs_sw_lt10}, but grouped by the preceding switching voltage rather than combined into a single recipe-level distribution. Across the measured voltage range, the median $|\Delta R_{\mathrm{vol}}|$ depends on the preceding switching voltage in all three recipe groups. Within the sampled bins, the lowest median values occur at $+2.0$~V for all three recipe groups, whereas the highest median values occur at $-3.0$~V. This behavior is therefore not symmetric between positive and negative bias.

HfO$_\text{x}$N$_\text{y}$ remains intermediate, with the median $|\Delta R_{\mathrm{vol}}|$ decreasing to 79.4~$\Omega$ at $+2.0$~V and increasing to 209.1~$\Omega$ at $-3.0$~V. TiO$_\text{x}$/AlO$_\text{y}$ remains the lowest-volatility recipe, with corresponding values of 3.77~$\Omega$ and 76.7~$\Omega$. TiO$_\text{x}$ shows the strongest volatility response and the strongest voltage dependence, with the median decreasing to 85.2~$\Omega$ at $+2.0$~V and increasing to 6281.5~$\Omega$ at $-3.0$~V. Taken together, these results show that volatility magnitude in the present dataset is both recipe-dependent and dependent on the preceding switching voltage, with the largest volatility observed under the most negative switching-voltage condition considered here.

\begin{figure}[!htbp]
    \centering
    \includegraphics[width=0.75\linewidth]{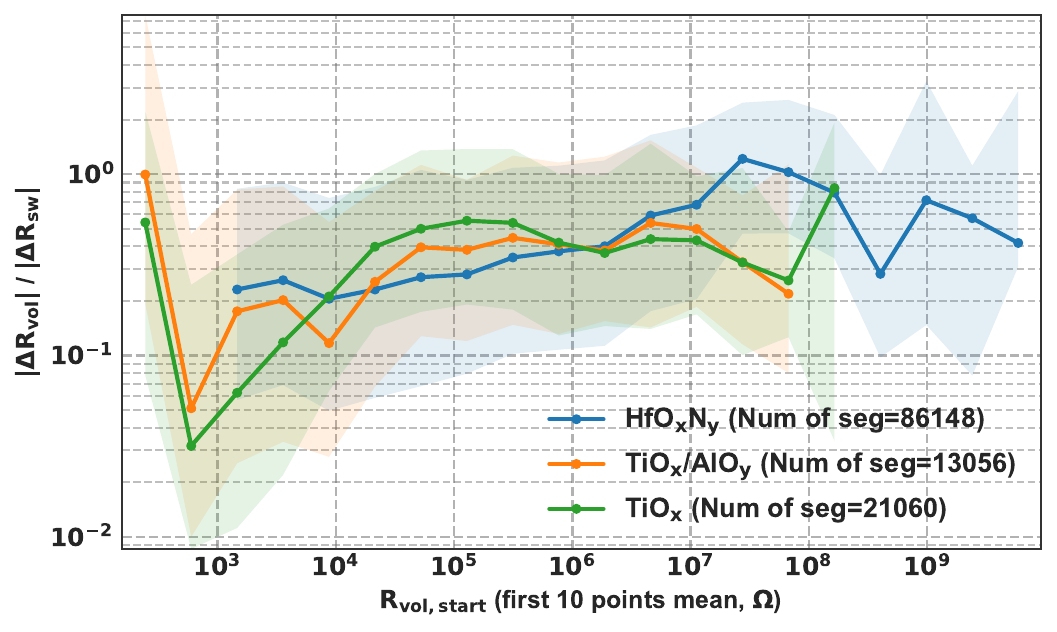}
    \caption{Ratio of absolute volatility magnitude to absolute switching magnitude, $|\Delta R_{\mathrm{vol}}|/|\Delta R_{\mathrm{sw}}|$, as a function of $R_{\mathrm{vol,start}}$ for the HfO$_\text{x}$N$_\text{y}$, TiO$_\text{x}$/AlO$_\text{y}$, and TiO$_\text{x}$ recipe groups, using the filtered PF-IR dataset. Markers show the median, shaded regions show the interquartile range, and the vertical axis is logarithmic.}
    \label{fig:vol_delta_abs_over_sw_delta_abs_vs_rstart_binned20_median_by_recipe_pubquality_abs10}
\end{figure}

Figure~\ref{fig:vol_delta_abs_over_sw_delta_abs_vs_rstart_binned20_median_by_recipe_pubquality_abs10} compares, for the three recipe groups, how large the volatility response is relative to the preceding switching response, using the same filtered PF-IR dataset as in the preceding volatility analysis. The plotted metric is the absolute ratio $|\Delta R_{\mathrm{vol}}|/|\Delta R_{\mathrm{sw}}|$, shown against the start-state resistance $R_{\mathrm{vol,start}}$ defined in Equation\,~\ref{eqn:volatility_delta}. In this way, the figure complements the absolute-volatility analysis by showing the fraction of the preceding switching change that remains as post-pulse relaxation across different start-resistance regimes.

The retained dataset contains 86,148 segments for HfO$_\text{x}$N$_\text{y}$, 13,056 for TiO$_\text{x}$/AlO$_\text{y}$, and 21,060 for TiO$_\text{x}$. The corresponding overall median values of $|\Delta R_{\mathrm{vol}}|/|\Delta R_{\mathrm{sw}}|$ are 0.2606, 0.2497, and 0.3283, respectively. Thus, for all three recipe groups, the typical volatility magnitude remains smaller than the corresponding switching magnitude, with TiO$_\text{x}$ showing the largest overall median ratio and HfO$_\text{x}$N$_\text{y}$ and TiO$_\text{x}$/AlO$_\text{y}$ remaining slightly lower and relatively similar.

The dependence on $R_{\mathrm{vol,start}}$ is nevertheless non-uniform. By visual inspection, the lowest-resistance bins show more elevated and variable ratios, whereas the intermediate start-state range becomes comparatively stable and remains broadly below unity. Within this intermediate range, HfO$_\text{x}$N$_\text{y}$ stays comparatively lower, TiO$_\text{x}$/AlO$_\text{y}$ occupies a similar but slightly higher level, and TiO$_\text{x}$ is generally highest. At higher start resistance, the recipe dependence becomes stronger again: HfO$_\text{x}$N$_\text{y}$ shows the most pronounced upward turn and reaches a maximum binned median ratio of 1.215, indicating that in some high-resistance bins the volatility magnitude can slightly exceed the preceding switching magnitude. TiO$_\text{x}$/AlO$_\text{y}$ approaches unity with a maximum binned median of 0.995, whereas TiO$_\text{x}$ reaches 0.837. Overall, the relative volatility response remains smaller than the preceding switching response in most of the sampled range, but can approach or slightly exceed unity in selected start-state regimes.

\clearpage
\begingroup
\scriptsize
\setlength{\tabcolsep}{4pt}
\renewcommand{\arraystretch}{1.08}
\setlength{\LTleft}{0pt}
\setlength{\LTright}{0pt}

\begin{longtable}{@{}>{\raggedleft\arraybackslash}p{0.07\linewidth}>{\raggedright\arraybackslash}p{0.44\linewidth}>{\raggedright\arraybackslash}p{0.44\linewidth}@{}}
\caption{Mapping between the numbered feature labels in Figure~\ref{fig:cct_feature_relationships}(a) and the corresponding database tables and columns.}
\label{tab:feature_mapping}\\
\toprule
\textbf{Display ID} & \textbf{Database table} & \textbf{Database column or derived quantity} \\
\midrule
\endfirsthead

\multicolumn{3}{c}{\tablename~\thetable\ continued}\\
\toprule
\textbf{Display ID} & \textbf{Database table} & \textbf{Database column or derived quantity} \\
\midrule
\endhead

\midrule
\multicolumn{3}{r}{Continued on next page}\\
\endfoot

\bottomrule
\endlastfoot

0 & \nolinkurl{Features_Electroforming_max_drop} & \nolinkurl{electroform_voltage_V} \\
1 & \nolinkurl{Features_Electroforming_sinh} & \nolinkurl{ef_class} \\
2 & \nolinkurl{Features_IV_nonlinearity_sinh} & \nolinkurl{pos_up_comp_hit} \\
3 & \nolinkurl{Features_IV_nonlinearity_sinh} & \nolinkurl{neg_up_comp_hit} \\
4 & \nolinkurl{Features_IV_nonlinearity_sinh} & \nolinkurl{pos_down_comp_hit} \\
5 & \nolinkurl{Features_IV_nonlinearity_sinh} & \nolinkurl{neg_down_comp_hit} \\
6 & \nolinkurl{Features_IV_nonlinearity_sinh} & \nolinkurl{pos_up_a} \\
7 & \nolinkurl{Features_IV_nonlinearity_sinh} & \nolinkurl{pos_down_a} \\
8 & \nolinkurl{Features_IV_nonlinearity_sinh} & \nolinkurl{neg_down_a} \\
9 & \nolinkurl{Features_IV_nonlinearity_sinh} & \nolinkurl{neg_up_a} \\
10 & \nolinkurl{Function_CurveTracer} & \nolinkurl{NEG(negative_voltage_max_V)} \\
11 & \nolinkurl{Features_IV_nonlinearity_sinh} & \nolinkurl{pos_down_r2_raw} \\
12 & \nolinkurl{Features_IV_nonlinearity_sinh} & \nolinkurl{neg_down_r2_raw} \\
13 & \nolinkurl{Features_IV_nonlinearity_sinh} & \nolinkurl{pos_up_r2_raw} \\
14 & \nolinkurl{Features_IV_nonlinearity_sinh} & \nolinkurl{neg_up_r2_raw} \\
15 & \nolinkurl{Subdie} & \nolinkurl{cross_sectional_area_um2} \\
16 & \nolinkurl{Features_Switching_Volatility_Delta_result} & \nolinkurl{sw_segment_voltage_V} \\
17 & \nolinkurl{Function_ParameterFit_interRetention} & \nolinkurl{pulse_width_us} \\
18 & \nolinkurl{Features_Switching_Volatility_Delta_result} & \nolinkurl{ABS(sw_segment_voltage_V)} \\
19 & \nolinkurl{Die} & \nolinkurl{die_number} \\
20 & \nolinkurl{Features_Switching_Volatility_Delta_result} & \nolinkurl{sw_delta_ohm} \\
21 & \nolinkurl{Features_Switching_Volatility_Delta_result} & \nolinkurl{vol_delta_ohm} \\
22 & \nolinkurl{Features_Switching_Volatility_Delta_result} & \nolinkurl{ABS(sw_delta_ohm)} \\
23 & \nolinkurl{Features_Switching_Volatility_Delta_result} & \nolinkurl{ABS(vol_delta_ohm)} \\
24 & \nolinkurl{Features_Switching_Volatility_Delta_result} & \nolinkurl{sw_start_stat_ohm} \\
25 & \nolinkurl{Features_Switching_Volatility_Delta_result} & \nolinkurl{sw_end_stat_ohm} \\
26 & \nolinkurl{Features_Switching_Volatility_Delta_result} & \nolinkurl{vol_start_stat_ohm} \\
27 & \nolinkurl{Features_Switching_Volatility_Delta_result} & \nolinkurl{vol_end_stat_ohm} \\
28 & \nolinkurl{Features_IV_nonlinearity_sinh} & \nolinkurl{neg_up_b} \\
29 & \nolinkurl{Features_IV_nonlinearity_sinh} & \nolinkurl{pos_up_b} \\
30 & \nolinkurl{Features_Ron_Roff_sinh} & \texttt{pos\_roff\_ohm / pos\_ron\_ohm} \\
31 & \nolinkurl{Features_Ron_Roff_sinh} & \texttt{neg\_roff\_ohm / neg\_ron\_ohm} \\
32 & \nolinkurl{Features_IV_nonlinearity_sinh} & \nolinkurl{neg_down_b} \\
33 & \nolinkurl{Features_IV_nonlinearity_sinh} & \nolinkurl{pos_down_b} \\
34 & \nolinkurl{Features_Ron_Roff_sinh} & \nolinkurl{pos_ron_ohm} \\
35 & \nolinkurl{Features_Ron_Roff_sinh} & \nolinkurl{neg_ron_ohm} \\
36 & \nolinkurl{Features_Ron_Roff_sinh} & \nolinkurl{pos_roff_ohm} \\
37 & \nolinkurl{Features_Ron_Roff_sinh} & \nolinkurl{neg_roff_ohm} \\
38 & \nolinkurl{Device} & \nolinkurl{wordline} \\
39 & \nolinkurl{Device} & \nolinkurl{bitline} \\
40 & \nolinkurl{Function_ParameterFit} & \nolinkurl{pulse_width_us} \\
41 & \nolinkurl{Features_Electroforming_max_drop} & \nolinkurl{r_before_ohm} \\
42 & \nolinkurl{Features_Electroforming_max_drop} & \nolinkurl{drop_ratio} \\
43 & \nolinkurl{Wafer} & \nolinkurl{recipe_id} \\
44 & \nolinkurl{Function_CurveTracer} & \nolinkurl{positive_voltage_max_V} \\
45 & \nolinkurl{Features_Electroforming_max_drop} & \nolinkurl{r_after_ohm} \\
\midrule
\multicolumn{3}{@{}p{0.95\linewidth}@{}}{\textit{Notes:} \texttt{NEG(x)} denotes $-x$; some negative-polarity values are stored as positive magnitudes. \texttt{ABS(x)} denotes the absolute value of $x$. A ratio written as \texttt{a / b} denotes $a$ divided by $b$.}\\
\end{longtable}
\endgroup

\clearpage

\section{Storage benchmarking}
\label{subsec:Storage benchmarking}

All storage benchmarks were performed on a HONOR DRB-P laptop running 64-bit Windows 11. The system was equipped with an Intel Core Ultra 9 285H processor with 16 cores and 16 logical processors, 32\,GB RAM, and a YMTC PC411-1TB-B 1\,TB NVMe SSD. The benchmark scripts were executed using the project's Python 3.10.11 virtual environment, linked against SQLite 3.40.1, with pandas 2.3.3 and NumPy 2.2.6.

\subsection{Common-source data generation}
\label{subsec:supp_common_source_generation}

To enable a controlled comparison between single-table and multi-table data structures, a common synthetic source dataset was first generated for each benchmark scale using a fixed-seed random workflow. 

The canonical source retained the full hierarchy of wafer, die, subdie, device, experiment, and point records, together with function-specific parameter tables for \texttt{CurveTracer} (\texttt{CT}), \texttt{ParameterFit} (\texttt{PF}), and a generic retention-type readout module (\texttt{RET}). In the deposited dataset, the corresponding programming-plus-retention workflow is implemented as \texttt{ParameterFit\_interRetention} (\texttt{PF-IR}), where each programming segment is followed by an interleaved read segment. Three dataset sizes were considered: \(10^6\), \(5\times10^6\), and \(10^7\) point-level records. This common source was then materialized into two alternative data structures for benchmarking: a fully denormalized single-table representation and a normalized multi-table hierarchy.

\subsection{Single-table and multi-table storage benchmarks}
\label{subsec:supp_storage_benchmarks}

As described in Section~\ref{subsec:supp_common_source_generation}, the common-source dataset was materialized into two data structures for benchmarking: a fully denormalized single-table representation and a normalized multi-table hierarchy. Here, data format refers to the storage format itself, namely CSV, TSV, JSON Lines, and indexed SQLite, whereas data structure refers to the organization of the same logical content in either single-table or multi-table form. In this context, the normalized hierarchy follows the relational database principle of separating repeated entities into linked tables to reduce redundancy, whereas the denormalized representation stores the joined logical content in a single flattened table for direct access \cite{Codd1970,Date2003,ElmasriNavathe2016}.

For both data structures, four data formats were compared: CSV, tab-delimited text (TSV), JSON Lines, and indexed SQLite. For each data format, on-disk footprint together with create, read, query, update, and delete times were recorded. The create step measured the time required to write the corresponding data structure into the target data format, whereas the read step measured the time required to load the stored data back into memory.

The query workload was defined consistently across the two data structures. Records were filtered by \texttt{func\_name} equal to CT, \texttt{wafer\_id} equal to 1 or 2, \texttt{die\_index} less than 50, \texttt{v\_form} between 1.7 and 2.3, and \texttt{ct\_v\_start} greater than 0, and were then grouped by \texttt{device\_id} to calculate point count, mean voltage, and mean current. In the single-table benchmark, these operations were applied directly to the denormalized table. In the multi-table benchmark, the same logic was applied after joining the experiment table with the device, subdie, die, wafer, \texttt{param\_ct}, and point tables.

Update and delete were both defined at the experiment level. For each dataset scale, 1\% of the unique experiment identifiers were sampled without replacement, and the same sampled subset was used for both operations. In both benchmarks, update increased \texttt{v\_form} by 0.1, \texttt{voltage} by 0.01, and the corresponding function-specific parameter field by 0.1. In the single-table benchmark, these changes were applied to rows in the denormalized table. In the multi-table benchmark, the same updates were distributed across the experiment, point, and function-specific parameter tables. Delete removed all records associated with the selected experiments; in the multi-table benchmark, this affected the point, parameter, and experiment tables, whereas hierarchy tables were retained. Update and delete were timed independently from the same original dataset state.

CSV, TSV, and JSON Lines were treated as plain-text formats, so query, update, and delete were carried out by loading the stored data into memory and writing the processed results back to disk. SQLite was treated as an indexed relational database format, allowing the same logical operations to be executed directly in the database. The main distinction between the two benchmarks, therefore, lies in the data structure: the single-table benchmark uses a flat denormalized structure, whereas the multi-table benchmark uses a normalized relational structure with linked tables.

\subsection{Storage benchmarking results}
\label{subsec:supp_storage_benchmark_results}

Figure~\ref{fig:storage_benchmark} from the main text summarizes the benchmark results for the common-source dataset under the fully denormalized single-table representation and the normalized multi-table hierarchy. At \(10^7\) point-level records, the single-table storage footprints were 2065\,MiB for CSV/TSV, 4863\,MiB for JSON Lines, and 1543\,MiB for indexed SQLite, whereas the corresponding multi-table footprints were 762\,MiB, 1146\,MiB, and 588\,MiB, respectively. (1 MiB = 1024$^2$ bytes, 1 GiB = 1024$^3$ bytes) This corresponds to footprint reductions of 63.1\% for CSV/TSV, 76.4\% for JSON Lines, and 61.9\% for SQLite in the normalized representation, consistent with the general expectation that normalized schemas reduce repeated metadata storage by separating shared attributes from high-volume fact records \cite{Codd1970,Date2003,ElmasriNavathe2016}.

Data-format-dependent differences were also clear. At \(10^7\) records in the single-table benchmark, query latencies were 13.8\,s for CSV, 17.6\,s for TSV, 428.0\,s for JSON Lines, and 0.678\,s for SQLite; the corresponding update latencies were 77.0\,s, 80.1\,s, 450.5\,s, and 1.86\,s; and delete latencies were 77.0\,s, 79.5\,s, 468.3\,s, and 0.152\,s. In the multi-table benchmark at the same scale, query latencies were 4.48\,s, 5.02\,s, 19.7\,s, and 0.216\,s; update latencies were 22.9\,s, 23.2\,s, 27.8\,s, and 0.055\,s; and delete latencies were 22.9\,s, 22.6\,s, 28.0\,s, and 0.060\,s, for CSV, TSV, JSON Lines, and SQLite, respectively. The consistently stronger query, update, and delete performance of indexed SQLite is consistent with the well-established advantages of indexed relational execution over repeated full-file parsing and rewriting \cite{Date2003,Silberschatz2020}.

These results separate two effects. First, data format primarily determined the absolute performance level across query, update, and delete operations: indexed SQLite was consistently the fastest format, whereas JSON Lines became the slowest at larger scales, especially for read-modify-rewrite workloads. CSV and TSV remained broadly similar to each other across all scales. Second, data structure primarily determined storage footprint and scaling behavior. At \(10^7\) records, moving from the fully denormalized single-table structure to the normalized multi-table hierarchy reduced query time by 3.09$\times$ for CSV, 3.51$\times$ for TSV, 21.8$\times$ for JSON Lines, and 3.13$\times$ for SQLite; update time by 3.36$\times$, 3.46$\times$, 16.2$\times$, and 33.8$\times$; and delete time by 3.36$\times$, 3.52$\times$, 16.7$\times$, and 2.53$\times$, respectively. This pattern is consistent with normalization reducing redundancy and thereby lowering the amount of data that must be scanned or rewritten during modification-heavy workloads \cite{Codd1970,ElmasriNavathe2016,Silberschatz2020}.

A small-scale exception was observed at \(10^6\) records, where the denormalized table could still be competitive or faster for some query cases since it avoided joins. For example, SQLite query latency was 0.016\,s in the single-table benchmark and 0.231\,s in the multi-table benchmark at this scale. Such behavior is also consistent with the familiar trade-off between normalization and denormalization: denormalized layouts can simplify access patterns and reduce join overhead, but they also amplify repeated attributes and rewrite volume as dataset size grows \cite{Date2003,KimballRoss2013}. Taken together, the benchmark distinguishes the effects of data format from those of data structure. Data format, represented here by CSV, TSV, JSON Lines, and indexed SQLite, primarily determines the absolute performance level. Data structure, represented here by the fully denormalized single-table layout and the normalized multi-table hierarchy, primarily determines storage footprint and scaling behavior.

The practical importance of these data formats and structures is also evident in the deployed database. \texttt{Experimental\_Detail} alone occupies 9.24\,GiB, with a further 2.04\,GiB from its main experiment-linked index, while storing only 8 point-level fields. A simple flat-schema heuristic, therefore, suggests that repeating broader schema context at the point level would push storage well into the tens-of-times range, even before accounting for additional index overhead or one-to-many and many-to-many expansion. Given the retained dataset size of 169,270,462 point-level rows, the indexed relational format and normalized multi-table structure are practical requirements rather than optional design choices.

\clearpage
\bibliographystyle{unsrt}
\bibliography{bib/SciData_Combined_bibliography}

\end{document}